\documentclass[twocolumn, twocolappendix]{aastex631}

\usepackage{newtxtext,newtxmath}
\usepackage[normalem]{ulem}
\usepackage[T1]{fontenc}

\usepackage{graphicx}	
\usepackage{amsmath}	
\usepackage{mathtools}
\usepackage{fontawesome}
\usepackage{hyperref}
\usepackage{xspace}
\usepackage{dsfont}
\usepackage{graphicx}
\usepackage{graphbox}
\usepackage{bbm}
\usepackage{soul}
\usepackage{float}
\usepackage{tabularx}
\usepackage{multirow}
\usepackage{braket}

\makeatletter
\let\frontmatter@title@above=\relax
\makeatother

\newcommand{\nn}{NeHOD\xspace}

\newcommand{\vect}[1]{\boldsymbol{#1}}
\newcommand{\zzt}[1]{\ensuremath{\mathbf{z}_{#1}}\xspace}
\newcommand{\uut}[1]{\ensuremath{\mathbf{u}_{#1}}\xspace}

\newcommand{\yy}{\ensuremath{\mathbf{y}}\xspace}
\newcommand{\zz}{\ensuremath{\mathbf{z}}\xspace}
\newcommand{\uu}{\ensuremath{\mathbf{u}}\xspace}
\newcommand{\vtheta}{\ensuremath{\vect{\theta}}\xspace}
\newcommand{\vthetasim}{\ensuremath{\vect{\theta}_\mathrm{sim}}\xspace}
\newcommand{\vthetasat}{\ensuremath{\vect{\theta}_\mathrm{sat}}\xspace}
\newcommand{\vthetahalo}{\ensuremath{\vect{\theta}_\mathrm{halo}}\xspace}
\newcommand{\vthetacent}{\ensuremath{\vect{\theta}_\mathrm{cent}}\xspace}

\newcommand{\msubhalo}{\ensuremath{M_{\mathrm{sub}}}\xspace}
\newcommand{\msubhalostar}{\ensuremath{M_\mathrm{sub, \star}}\xspace}
\newcommand{\msubhaloi}{\ensuremath{M_{\mathrm{sub}, i}}\xspace}
\newcommand{\msubhalostari}{\ensuremath{M_{\mathrm{sub, \star}, i}}\xspace}

\newcommand{\mhalo}{\ensuremath{M_\mathrm{halo}}\xspace}
\newcommand{\mstar}{\ensuremath{M_\mathrm{\star}}\xspace}
\newcommand{\mcentstar}{\ensuremath{M_\mathrm{cent, \star}}\xspace}
\newcommand{\mcent}{\ensuremath{M_\mathrm{cent}}\xspace}
\newcommand{\nsat}{\ensuremath{N_\mathrm{sat}}\xspace}
\newcommand{\vmaxtilde}{\ensuremath{\widetilde{V}_\mathrm{max}}\xspace}

\newcommand{\vmaxtildei}{\ensuremath{\widetilde{V}_{\mathrm{max}, i}}\xspace}

\newcommand{\mwdm}{\ensuremath{m_\mathrm{WDM}}\xspace}
\newcommand{\imwdm}{\ensuremath{m^{-1}_\mathrm{WDM}}\xspace}
\newcommand{\sno}{\ensuremath{A_\mathrm{SN1}}\xspace}
\newcommand{\snt}{\ensuremath{A_\mathrm{SN2}}\xspace}
\newcommand{\agn}{\ensuremath{A_\mathrm{AGN}}\xspace}
\newcommand{\simparam}{\ensuremath{\{\mwdm, \sno, \snt, \agn\}}\xspace}
\newcommand{\astroparam}{\ensuremath{\{\sno, \snt, \agn\}}\xspace}

\newcommand{\kpch}{\ensuremath{\mathrm{kpc}\,h^{-1}}\xspace}

\newcommand{\modoth}{\ensuremath{\mathrm{M_\odot} h^{-1}}\xspace}

\defcitealias{cosmovdm}{CM23}

\begin{document}

\title[Emulating DREAMS with Variational Diffusion Models]{How DREAMS are made: Emulating Satellite Galaxy and Subhalo Populations with Diffusion Models and Point Clouds}

\correspondingauthor{Tri Nguyen} \\
\email{tnguy@mit.edu}

\author[0000-0001-6189-8457]{Tri Nguyen}
\affiliation{Department of Physics, Massachusetts Institute of Technology, Cambridge, MA 02139, USA}
\affiliation{Kavli Institute for Astrophysics and Space Research, Massachusetts Institute of Technology, Cambridge, MA 02139, USA}
\affiliation{The NSF AI Institute for Artificial Intelligence and Fundamental Interactions, Cambridge, MA 02139, USA}
\affiliation{Department of Physics and Astronomy and CIERA, Northwestern University, 2145 Sheridan Road, Evanston, IL 60208}

\author[0000-0002-4816-0455]{Francisco Villaescusa-Navarro}
\affiliation{Center for Computational Astrophysics, Flatiron Institute, 162 5th Avenue, New York, NY 10010, USA}
\affiliation{Department of Astrophysical Sciences, Princeton University, Peyton Hall, Princeton, NJ 08544, USA}

\author[0000-0001-9088-7845]{Siddharth Mishra-Sharma}
\altaffiliation{Currently at Anthropic; worked performed while at MIT/IAIFI.}
\affiliation{Department of Physics, Massachusetts Institute of Technology, Cambridge, MA 02139, USA}
\affiliation{The NSF AI Institute for Artificial Intelligence and Fundamental Interactions, Cambridge, MA 02139, USA}
\affiliation{Center for Theoretical Physics, Massachusetts Institute of Technology, Cambridge, MA 02139, USA}
\affiliation{Department of Physics, Harvard University, Cambridge, MA 02138, USA}

\author[0000-0001-9088-7845]{Carolina Cuesta-Lazaro}
\affiliation{Department of Physics, Massachusetts Institute of Technology, Cambridge, MA 02139, USA}
\affiliation{The NSF AI Institute for Artificial Intelligence and Fundamental Interactions, Cambridge, MA 02139, USA}
\affiliation{Center for Astrophysics, Harvard \& Smithsonian, Cambridge, MA 02138, USA}

\author[0000-0002-5653-0786]{Paul Torrey}
\affiliation{Department of Astronomy, University of Virginia, 530 McCormick Road, Charlottesville, VA 22904 USA}

\author[0000-0003-0777-4618]{Arya~Farahi}
\affiliation{Departments of Statistics and Data Science, University of Texas at Austin, Austin, TX 78757, USA}

\author[0000-0002-8111-9884]{Alex M. Garcia}
\affiliation{Department of Astronomy, University of Virginia, 530 McCormick Road, Charlottesville, VA 22904 USA}

\author[0000-0002-2628-0237]{Jonah C. Rose}
\affiliation{Center for Computational Astrophysics, Flatiron Institute, 162 5th Avenue, New York, NY 10010, USA}
\affiliation{Department of Astronomy, University of Florida, Gainesville, FL 32611, USA}

\author[0000-0002-7968-2088]{Stephanie O'Neil}
\affiliation{Department of Physics, Massachusetts Institute of Technology, Cambridge, MA 02139, USA}
\affiliation{Kavli Institute for Astrophysics and Space Research, Massachusetts Institute of Technology, Cambridge, MA 02139, USA}

\author[0000-0001-8593-7692]{Mark Vogelsberger}
\affiliation{Department of Physics, Massachusetts Institute of Technology, Cambridge, MA 02139, USA}
\affiliation{Kavli Institute for Astrophysics and Space Research, Massachusetts Institute of Technology, Cambridge, MA 02139, USA}
\affiliation{The NSF AI Institute for Artificial Intelligence and Fundamental Interactions, Cambridge, MA 02139, USA}

\author[0000-0002-6196-823X]{Xuejian Shen}
\affiliation{Department of Physics, Massachusetts Institute of Technology, Cambridge, MA 02139, USA}
\affiliation{Kavli Institute for Astrophysics and Space Research, Massachusetts Institute of Technology, Cambridge, MA 02139, USA}

\author[0000-0002-3400-6991]{Cian Roche}
\affiliation{Department of Physics, Massachusetts Institute of Technology, Cambridge, MA 02139, USA}
\affiliation{Kavli Institute for Astrophysics and Space Research, Massachusetts Institute of Technology, Cambridge, MA 02139, USA}

\author[0000-0001-5769-4945]{Daniel Angl\'es-Alc\'azar}
\affiliation{Department of Physics, University of Connecticut, 196 Auditorium Road, U-3046, Storrs, CT 06269-3046, USA}
\affiliation{Center for Computational Astrophysics, Flatiron Institute, 162 5th Avenue, New York, NY 10010, USA}

\author[0000-0002-3204-1742]{Nitya Kallivayalil}
\affiliation{Department of Astronomy, University of Virginia, 530 McCormick Road, Charlottesville, VA 22904 USA}

\author[0000-0002-8984-0465]{Julian B.~Mu\~noz}
\affiliation{Department of Astronomy, University of Texas at Austin, Austin, TX 78757, USA}

\author[0000-0002-7939-2988]{Francis-Yan Cyr-Racine}
\affiliation{Department of Physics and Astronomy, University of New Mexico, 210 Yale Blvd NE, Albuquerque, NM 87106, USA}

\author[0000-0002-7638-7454]{Sandip Roy}
\affiliation{Department of Physics, Princeton University, Princeton, NJ 08544, USA}

\author[0000-0003-2806-1414]{Lina Necib}
\affiliation{Department of Physics, Massachusetts Institute of Technology, Cambridge, MA 02139, USA}
\affiliation{Kavli Institute for Astrophysics and Space Research, Massachusetts Institute of Technology, Cambridge, MA 02139, USA}
\affiliation{The NSF AI Institute for Artificial Intelligence and Fundamental Interactions, Cambridge, MA 02139, USA}

\author[0000-0003-4004-2451]{Kassidy E. Kollmann}
\affiliation{Department of Physics, Princeton University, Princeton, NJ 08544, USA}

\begin{abstract}

The connection between galaxies and their host dark matter (DM) halos is critical to our understanding of cosmology, galaxy formation, and DM physics. 
To maximize the return of upcoming cosmological surveys, we need an accurate way to model this complex relationship.
Many techniques have been developed to model this connection, from Halo Occupation Distribution (HOD) to empirical and semi-analytic models to hydrodynamic. 
Hydrodynamic simulations can incorporate more detailed astrophysical processes but are computationally expensive; HODs, on the other hand, are computationally cheap but have limited accuracy.
In this work, we present \nn, a generative framework based on variational diffusion model and Transformer, for painting galaxies/subhalos on top of DM with an accuracy of hydrodynamic simulations but at a computational cost similar to HOD.  
By modeling galaxies/subhalos as point clouds, instead of binning or voxelization, we can resolve small spatial scales down to the resolution of the simulations.
For each halo, \nn predicts the positions, velocities, masses, and concentrations of its central and satellite galaxies. 
We train \nn on the TNG-Warm DM suite of the DREAMS project, which consists of 1024 high-resolution zoom-in hydrodynamic simulations of Milky Way-mass halos with varying warm DM mass and astrophysical parameters. 
We show that our model captures the complex relationships between subhalo properties as a function of the simulation parameters, including the mass functions, stellar-halo mass relations, concentration-mass relations, and spatial clustering.
Our method can be used for a large variety of downstream applications, from galaxy clustering to strong lensing studies.
\end{abstract}


\section{Introduction}
\label{section:intro}

The abundance and spatial clustering of galaxies and dark matter (DM) halos is sensitive to assumptions about the underlying galaxy formation physics and cosmological model, including the nature of DM. 
Warm DM (WDM) models, for example, will predict fewer massive subhalos compared to cold DM due to free-streaming in the early Universe, thereby affecting observable properties such as the satellite abundance, their luminosity functions, and spatial distribution~\cite[e.g.,][]{2001ApJ...556...93B, 2014MNRAS.439..300L, 2016MNRAS.455..318B}.
Likewise, baryonic processes, including stellar winds, feedback from supernovae (SN) and active galactic nuclei (AGN), can disrupt the formation of less massive subhalos and redistribute matter within halos, thus altering galaxy population properties in similar ways~\cite[e.g.,][]{2010Governato, 2011Teyssier, 2017Peirani, 2019Peirani, 2021EnglerA, 2023Gebhardt}. 
The properties of galaxies and their distributions thus serve as powerful probes for DM models and galaxy formation physics.

To leverage these properties, several current surveys are designed to measure these cosmic structures, ranging from large-scale structures to deep imaging of nearby galaxies. 
Surveys such as the Sloan Digital Sky Survey~\cite[SDSS;][]{2000AJ....120.1579Y}, Dark Energy Survey~\cite[DES;][]{2005astro.ph.10346T, 2018ApJS..239...18A, 2021ApJS..255...20A}, Dark Energy Spectroscopic Instrument~\cite[DESI;][]{2016arXiv161100036D}, and Euclid~\citep{2022A&A...662A.112E} map the 3-dimensional distributions and clustering of billions of galaxies.
Surveys such as Satellites Around Galactic Analogs~\cite[SAGA;][]{2017ApJ...847....4G, 2021ApJ...907...85M, 2024arXiv240414498M, 2024arXiv240414499G, 2024arXiv240414500W} and Exploring the Local Volume in the Extended Solar neighborhood~\cite[ELVES;][]{2022ApJ...933...47C} have characterized hundreds of satellite dwarf galaxies around the Milky Way and nearby galaxies. 
In addition, upcoming surveys such as the Rubin Observatory's Legacy Survey of Space and Time~\citep[LSST;][]{LSST:2019} and Roman Space Telescope~\citep{2019arXiv190205569A} will have the capacity to look deeper and wider, detecting fainter galaxies and providing a more complete sample of galaxies and their distributions. 
To maximize the information we can extract from these surveys (through, e.g., galaxy clustering or strong lensing studies) and obtain robust constraints on cosmology, astrophysics, and DM properties, we must develop accurate theoretical predictions for galaxies/subhalos as a function of such properties. 

Hydrodynamic simulations are the primary method in modern astrophysics for generating theoretical predictions about the phase-space distribution of galaxies and their intrinsic properties (e.g., stellar mass, metallicity, neutral hydrogen mass, etc.).
These simulations model the evolution of galaxies and DM halos by solving fluid dynamics and gravitational equations within an expanding universe, starting from the initial density fluctuations derived from measurements of the Cosmic Microwave Background~\cite[for a review, see][]{2020NatRP...2...42V}.
Hydrodynamic simulations incorporate detailed prescriptions of baryonic processes (e.g., gas cooling, star formation, stellar/AGN feedback, etc.) and thus have been instrumental in understanding galaxies and their DM halos. 
Key predictions of these simulations range from the large-scale distribution of DM and galaxies~\cite[e.g.,][]{2006Natur.440.1137S, 2015MNRAS.446..521S, 2018MNRAS.475..676S, 2018MNRAS.475..648P, 2018MNRAS.480.5113M}, to the properties of the intergalactic and interstellar medium~\cite[e.g.,][]{1996ApJ...457L..51H, 2014ApJ...793...30G, 2016MNRAS.463.1462O, 2018MNRAS.479..994R, 2018MNRAS.477.1206N, 2022MNRAS.511.4005K, 2022MNRAS.512.3243S}, down to the internal structure of DM halos~\cite[e.g.,][]{1996ApJ...462..563N, 1996MNRAS.283L..72N}, the properties of individual galaxies~\cite[e.g.,][]{2014Natur.509..177V, 2015MNRAS.454...83W, 2017MNRAS.467.4739K, 2017MNRAS.467..179G, 2018MNRAS.480..800H, 2018MNRAS.475..624N}, and the populations of satellite galaxies~\cite[e.g.,][]{2016ApJ...827L..23W, 2017MNRAS.471.1709G, 2017MNRAS.471.3547F, 2018PhRvL.121u1302K, 2018MNRAS.478..548S, 2019MNRAS.483.1314B, 2022NatAs...6..897S}.

To deepen our understanding of the physics governing DM and galaxy formation, we want to thoroughly explore how observed galaxy properties (e.g., abundance and spatial distribution) vary across different DM and astrophysical parameters. 
Ideally, this would involve conducting a large set of large-volume, high-resolution hydrodynamic simulations that cover this extensive parameter space.
However, hydrodynamic simulations are computationally demanding, which limits the simulated volume and/or resolution.
Large-volume simulations such as IllustrisTNG~\citep{2018MNRAS.473.4077P, 2017MNRAS.465.3291W} and MilleniumTNG~\citep{2005Natur.435..629S, 2023MNRAS.524.2556H} are constrained by mass resolution and cannot adequately capture the properties of low-mass and/or high-redshift galaxies. 
Conversely, zoom-in simulations such as FIRE~\citep{2014MNRAS.445..581H}, NIHAO~\citep{2015MNRAS.454...83W}, APOSTLE~\citep{2016MNRAS.457.1931S}, and Auriga~\citep{2017MNRAS.467..179G} provide higher resolution but have smaller volumes, resulting in a more limited number of simulated galaxies.
In addition, zoom-in simulations have difficulty modeling very massive halos given the smaller box sizes.

These challenges demonstrate the need for developing more efficient methods to study the relationship between galaxies and their DM halos.
A common way is to map galaxy properties onto DM halos simulated in N-body (i.e., gravity-only) simulations, which have a much lower computational cost than their full hydrodynamic counterparts.
For example, semi-analytic models, which use simplified baryonic physics prescriptions, and empirical models, which use empirical relationships from observational data (e.g., stellar-to-halo mass relations), can be used to assign galaxy properties to DM halos \citep[see][for a review]{Risa_2018}.

In this paper, we focus on Halo Occupation Distribution (HOD) methods, in which galaxies are assigned to DM halos in N-body simulations based on a statistical framework that describes the probability of a halo hosting a certain number of galaxies~\citep{1952ApJ...116..144N}.
This process, known as ``painting'' galaxies onto DM halos, follows a set of recipes that aim to capture the main physics describing the halo-galaxy connection.
HODs have been applied in various contexts, from studying galaxy clustering~\citep{2000MNRAS.311..793B, 2000MNRAS.318.1144P, 2016MNRAS.460.2552H} and the halo-galaxy connection~\citep{ 2003MNRAS.339.1057Y, 2009ApJ...696..620C, 2009MNRAS.394.1109C, Risa_2018}, to generating mock galaxy catalogs for recent surveys such as DESI~\citep{2024MNRAS.532..903S} and Euclid~\citep{2022A&A...658A..20H}. 
Despite these successes, since HOD methods rely on simplified assumptions about the halo-galaxy connection and are designed to match specific summary statistics (e.g., galaxy abundance and 2-point correlation function), they may fail for other summary statistics or probes of the entire observed distribution (e.g., field-level analyses)~\cite[e.g.,][]{2021JCAP...05..059S, 2021MNRAS.505.1422K, 2022JCAP...02..002W, 2023MNRAS.524.2507H}.

Ultimately, we want the accuracy of full hydrodynamic simulations at the computational cost of HODs.
In principle, machine learning models can reproduce and extend the work of traditional HODs, by not only learning the connection between halos and their constituent galaxies but also linking halo occupation to the underlying assumed physics.
Generative models, also called emulators, are particularly well-suited for this task as they learn directly from the outputs of simulations and generate new data samples, bypassing the need for summary statistics.
Normalizing flows~\citep{2015arXiv150505770J, 10.5555/3294771.3294994, 2019arXiv191202762P}, for example, have been used for HOD modeling by either painting galaxy properties on top of subhalos or modeling the position and velocity offsets between subhalos and galaxies~\citep[see e.g.,][]{Lovell_2023, Kwon_2024}.
Recent developments in diffusion models~\citep{2020arXiv200611239H, 2020arXiv201113456S, 2021arXiv210700630K} open up promising avenues for exploration, as these models are typically more expressive than flow-based models. 
In recent work, \cite{cosmovdm}, hereafter \citetalias{cosmovdm}, employs diffusion and point clouds to generate large-scale distributions of DM halos as a function of cosmological parameters.
Another recent work, \cite{2024arXiv240800839B}, combines convolutional neural networks and diffusion to predict the distributions of galaxy counts from DM fields and cosmological parameters, and demonstrates that such an approach outperforms HOD modeling.

In this paper, we explore how diffusion models and point clouds can be used to generate galaxies/subhalos within a single DM halo at field level, across different DM properties and baryonic physics scenarios.
We introduce the \nn framework, short for Neural Halo Occupation Distribution.
Distinct from most HODs and previous emulators, \nn employs a point-cloud approach to represent satellite galaxies (instead of relying on binning or voxelization), which allows it to resolve small spatial scales down to the resolution of the simulations.
In addition, \nn can take into account intra-galaxy (i.e., properties of a single galaxy) as well as inter-galaxy (i.e., between galaxies in a catalog) properties, since each point cloud/satellite galaxy population is generated at once using a Variational Diffusion Model with a Transformer-based noise prediction model~\citep{vaswani2023attention}.
The primary goal of \nn is to generate full galaxy catalogs, including the positions, velocities, and internal properties (such as mass and concentration) of halos, central galaxies, and satellite galaxies, given a set of simulation parameters. 
Thus, to complement the diffusion model, \nn uses normalizing flows~\citep{2015arXiv150505770J, 10.5555/3294771.3294994, 2019arXiv191202762P} to characterize halos and central galaxies.

We train our model using data from the DREAMS project~\citep{dreams}, which contains high-resolution, zoom-in simulations of Milky Way-mass halos.
Specifically, we utilize the TNG-Warm Dark Matter (WDM) zoom-in suite of simulations, in which each simulation varies over the astrophysical parameters (the strength of SN and AGN feedback) and the mass of the WDM particles.
When WDM decouples at early times, its free-streaming velocity is relativistic. 
These high speeds allow WDM particles to escape small potential wells, thus suppressing small-scale power~\citep{2000PhRvD..62f3511H, 2001ApJ...551..608S}. 
Consequently, WDM can influence properties of both halos and galaxies, such as the number of satellite galaxies, the satellite mass function, etc.~\citep{2001ApJ...556...93B, 2014MNRAS.439..300L, 2016MNRAS.455..318B, 2016MNRAS.461...60L, 2020MNRAS.498..702L}.
Additionally, WDM can affect the internal properties of galaxies~\citep[e.g.,][]{2010MNRAS.404L..16M, 2012MNRAS.420.2318L}, but only at sub-kpc scales for allowed WDM particle masses in the TNG-WDM suite.

By employing the TNG-WDM suite, we aim to capture variations due to both astrophysical processes and DM physics, as well as halo-to-halo variance, and demonstrate \nn's capability to model the complex relationships between halos, galaxies, and these parameters.
We highlight that emulating halos and galaxies under alternative DM models is, by itself, novel and interesting for understanding the impact of DM on galaxy formation at small scales and placing constraints on its properties.
However, we will explore these implications in future work, as this paper focuses primarily on the methodology and validation of our model.

Once trained, \nn can generate a catalog that represents galaxies residing in a halo, given some values of the astrophysical parameters and the WDM mass. 
Our model returns a galaxy catalog with the position, velocity, and internal properties of each galaxy inside the halo. 
We use multiple summary statistics to quantify the accuracy and precision of the generated galaxy catalogs and compare them with the ones from the simulations. 
We find \nn exhibits the accuracy of the full hydrodynamic simulations while being orders of magnitude faster.
This approach is particularly valuable in any context where populating halos with galaxies is necessary, ranging from galaxy redshift and lensing surveys~\citep{2003ApJ...593....1B, 2011ApJ...736...59Z, 2022A&A...658A..20H, 2024MNRAS.532..903S, 2024arXiv240718912C}, to near-field cosmology~\citep{2021MNRAS.502..621J, streamgen}, to high-redshift galaxies~\citep{2018MNRAS.480.3177B, 2012MNRAS.424.1363K, 2024MNRAS.529.3877R}.

This paper is structured as follows.
In \S\ref{section:sim}, we briefly summarize the DREAMS project and the WDM zoom-in suite of simulations used as the training dataset.
In \S\ref{section:method}, we describe the training dataset (in \S\ref{section:dataset}), the machine learning models (in \S\ref{section:vdm}), and the training procedure (in \S\ref{section:training}).
In \S\ref{section:result}, we show examples of generated halos and galaxies and quantify the fidelity of the trained emulator model.
In \S\ref{section:discussion}, we compare \nn to traditional modeling techniques and other emulator frameworks, and discuss its current limitations and future prospects.
Finally, we state our conclusions in \S\ref{section:conclusion}.

\section{Simulations}
\label{section:sim}

\subsection{The DREAMS project}
Introduced in~\cite{dreams}, the DREAMS project\footnote{\url{https://www.dreams-project.org/}} seeks to understand how cosmology, including DM models and baryonic physics, specifically feedback from SN and AGN, influence galaxy formation and evolution, ranging from galaxy clusters to ultra-faint satellites. 
Prescriptions for SN and AGN feedback are currently among the key sources of systematic uncertainty in cosmological simulations.
These feedback mechanisms present significant challenges in distinguishing the effects of baryonic processes from the inherent properties of DM, especially at the Galactic scales, e.g., by altering the inner density profiles in dwarf galaxies (\citealt{1996MNRAS.283L..72N, 2005AJ....129.2119S}; see also~\citealt{2017ARA&A..55..343B} for a review). 
Ultimately, DREAMS will help understand and disentangle these complex interactions.

The DREAMS suite comprises thousands of state-of-the-art hydrodynamic simulations and is the largest suite of hydrodynamic simulations that vary the DM physics. 
Similar to the CAMELS project~\citep{2021ApJ...915...71V}, these simulations vary both cosmological and astrophysical parameters.
Notably, these simulations also explore variations in DM particle properties, such as warm DM with different assumed particle masses.
The DREAMS suite includes both uniform box cosmological simulations, which provide extensive samples of halos and galaxies, as well as zoom-in cosmological simulations of Milky Way-mass hosts for higher-resolution studies. 
Among the simulations currently completed are the WDM Uniform Box and Milky Way suites.
For the full specifications of the simulations, we refer readers to Table 1 of \cite{dreams}.

Here, we briefly describe the TNG-WDM Milky Way suite used for the machine learning study in this work.
The TNG-WDM Milky Way suite contains 1024 simulations of Milky Way-mass halos with varying WDM mass and astrophysical parameters. 
These simulations are run with the moving-mesh code \textsc{Arepo}~\citep{2010MNRAS.401..791S, 2020ApJS..248...32W} which accounts for gravity and magneto-hydrodynamics, as well as a range of galaxy formation physics as included in the IllustrisTNG galaxy formation model.
The gravitational softening length at $z=0$ is $305 \, \mathrm{pc} \, h^{-1}$.
Cosmological parameters are fixed at $\Omega_\mathrm{m} = 0.301$, $\Omega_\Lambda = 0.698$, $\Omega_\mathrm{b} = 0.046$, $\sigma_8 = 0.839$, $h=0.691$, which is consistent with the Planck 2016 cosmology~\citep{2016A&A...594A..13P}.
The DM and baryon mass resolution are $1.2 \times 10^6\, \modoth$ and $1.9 \times 10^5 \, \modoth$ respectively, which is comparable to the mass resolution of the TNG50 simulation~\citep{2019MNRAS.490.3234N, 2019MNRAS.490.3196P}, which has a DM and baryon mass resolution of $6.5 \times 10^5\, \modoth$ and $1.2 \times 10^5 \, \modoth$. 

The simulations adopt the IllustrisTNG galaxy formation model, as detailed in \cite{2017MNRAS.465.3291W, 2018MNRAS.473.4077P}, which employs a tree and particle mesh (tree-PM) algorithm using periodic boundary conditions to simulate gravity.
This model also incorporates baryonic physics, such as AGN feedback, self-consistent magnetohydrodynamics, and an updated prescription of stellar formation and evolution, galactic winds, and outflows from the previous Illustris model~\citep{2014VogelsbergerA, 2014Torrey}.
The TNG-WDM suite adopts all of the fiducial TNG model parameters, except three parameters that control (i) SN feedback strength, (ii) SN-driven wind velocity, and (iii) AGN feedback strength. 
For a more detailed discussion of the baryonic prescription in DREAMS, we refer readers to \S 2.1 of~\cite{dreams}.
Here, we provide a brief description of the varied simulation parameters, including the WDM mass \mwdm and the three feedback parameters \astroparam, and the prior ranged in Table~\ref{tab:prior}.
The parameters are distributed uniformly in inverse WDM mass and log uniformly for the feedback parameters. 
These parameters are varied according to a Sobol sequence~\citep{sobol}, which ensures a quasi-random, well-distributed sampling across the parameter space that is suitable for machine learning applications.

\begin{table*}[hbt]
    \centering
    \begin{tabular}{lllll}
        \hline
         Parameter &  Description & Range &  Prior & TNG Fiducial\\ 
         \hline
         \mwdm  & Mass of WDM particles in keV & $[1.8,  30]$ & Inverse Uniform & $\infty$ \\
         \sno & Specific wind energy of SN wind & $[0.9, 14.4 ]$ &  Log Uniform & 3.6\\
         \snt & Dimensionless scaling factor of SN wind velocity & $[3.7, 14.8]$ &  Log Uniform & 7.4\\
         \agn & Fraction of energy transferred to nearby gas & $[0.025, 0.4] $ &  Log Uniform & 0.1\\
         & due to AGN accretion &  &  & \\
        \hline
    \end{tabular}
    \caption{Descriptions, prior ranges, and their TNG fiducial values of parameters of the 1024 simulations in TNG-WDM suite used in this work.}.
    \label{tab:prior}
\end{table*}

\subsection{Initial conditions and Milky Way selection criteria}
\label{section:sim_ic}

Each simulation zooms in on a high-resolution region that contains a Milky Way-mass halo.
The zoom-in procedure includes 3 steps: (1) simulate a low-resolution, DM-only uniform box, (2) randomly select a Milky Way-mass halo and perform a zoom-in DM-only simulation at an intermediate resolution, (3) follow with a final zoom-in with baryons at the fiducial resolution.
More details of the zoom-in procedure are outlined in Appendix A of \cite{dreams}.
Below, we highlight a few key points. 

We select the target halo randomly from the uniform box, with virial mass in the range of $(1.09-1.11) \times 10^{12} \, \modoth$.\footnote{This corresponds to $(1.58-1.61) \times 10^{12} \, \mathrm{M_\odot}$.} 
This is toward the upper end of the observed Milk Way-mass estimates in~\cite{2016ARA&A..54..529B}.
Additionally, the halo is selected such that it is not in the vicinity of another massive halo (with virial mass greater than $7.2 \times 10^{11} \modoth$) to avoid the high computational costs associated with simulating complex interactions in crowded regions.
As a result, the halos used in this work are not direct Local Group-analogs.

Initial conditions are generated by the MUlti-Scale Initial Conditions~\cite[MUSIC;][]{2011MNRAS.415.2101H} code at $z = 127$ using the second-order Lagrangian perturbation theory (2LPT).
The initial conditions for the final hydrodynamic zoom-in simulations are derived from intermediate-resolution, DM-only simulations. 
It is important to note that for each set of simulation parameters (e.g., the WDM mass and astrophysical parameters), a different uniform box was used, and a random target halo was selected from within this box. 
\textit{This approach ensures a diversity of initial conditions, so each halo in the training dataset represents a unique realization of a Milky Way-mass galaxy.}

Lastly, the halos and subhalos are identified using the friends-of-friends (\textsc{FoF}) and \textsc{subfind} algorithms~\citep{2001MNRAS.328..726S} with a linking length of $b=0.2$.
The \textsc{FoF} algorithm identifies halos by linking nearby DM particles into groups based on their spatial proximity.
\textsc{subfind} then refines these groups by identifying gravitationally bound substructures (subhalos) within the \textsc{FoF} halos.
In this work, halo properties are derived from \textsc{FoF}, while subhalo and galaxy properties (whether central or satellite) are from \textsc{subfind}.
We will explore the impact of the choice of the halo finder on the emulated halo and subhalo properties in future work.
We only include subhalos with at least 100 DM particles, which corresponds to a minimum mass of $1.2 \times 10^8 \, \modoth$; below this threshold, \textsc{subfind} does not reliably identify subhalos.
Additionally, we only consider subhalos that host galaxies, with stellar mass greater than $2 \times 10^5 \, \modoth$,  thus excluding dark subhalos.

\section{Methodology}
\label{section:method}

We begin with a brief overview of the \nn framework in \S\ref{section:method_overview}.
In \S\ref{section:dataset}, we describe the procedure for creating and preprocessing the dataset as well as the halo properties used as input and output features. 
We then describe the \nn architecture and training strategy in \S\ref{section:vdm} and \ref{section:training}, respectively.

\subsection{Overview of the \nn framework}
\label{section:method_overview}

The \nn (Neural Halo Occupation Distribution) framework consists of two main components: 
\begin{enumerate}
    \item A normalizing flow that models the conditional likelihood of halos and the galaxies residing at their center (i.e., the central galaxies), including properties such as total mass, stellar mass, and DM concentrations, given the simulation parameters:
    \begin{equation}    
    p(\vthetahalo, \vthetacent | \vthetasim),
    \end{equation}
    where \vthetahalo, \vthetacent, and \vthetasim are the halo, central galaxy, and simulation parameters respectively. 

    \item A Variational Diffusion Model (VDM) that models the conditional likelihood of satellite galaxies (e.g., positions, velocities, and internal properties), given the halo, central galaxy, and simulation parameters: 
    \begin{equation}
    p(\vthetasat | \vthetahalo, \vthetacent, \vthetasim),
    \end{equation}
    where \vthetasat are the properties of \textit{all} satellites within the halo.
\end{enumerate}

\begin{figure*}
    \centering
    \includegraphics[width=0.95\linewidth]{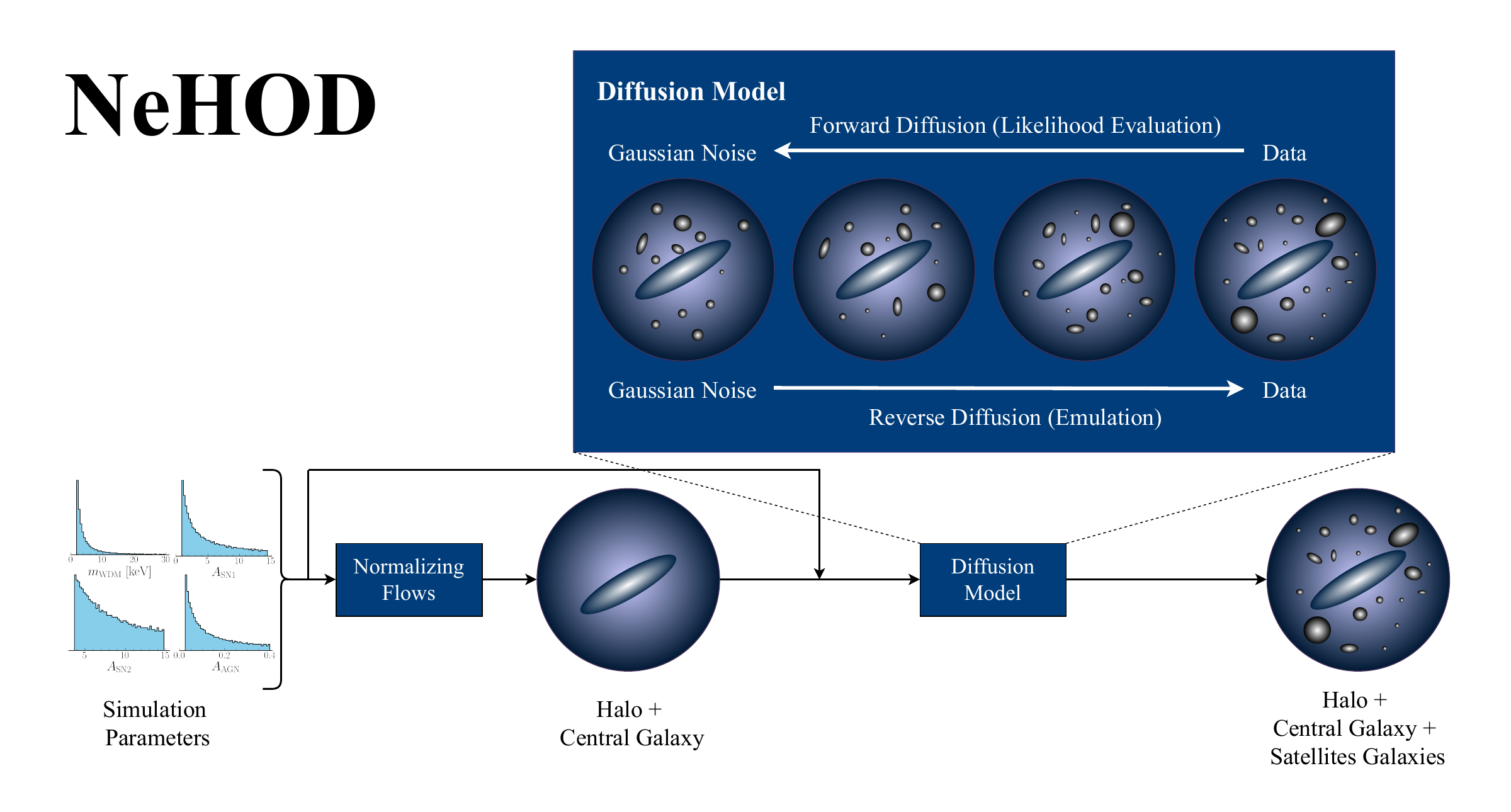}
    \caption{
    Flow chart of the \nn framework. 
    Black arrows indicate the flow of information into and out of the models. 
    During inference, simulation parameters (i.e., WDM mass and astrophysical parameters) are first input as conditioning (context) features into the normalizing flows, which then generate a halo and a central galaxy. 
    The properties of the halo and central galaxy, along with the simulation parameters, are subsequently passed as conditioning features into the VDM to generate satellite galaxies. 
    During training, the normalizing flows and the VDM are optimized independently (see Appendix~\ref{app:flows} and \ref{app:vdm} for details on the optimization objectives).
    }
    \label{fig:flow_chart}
\end{figure*}

The models operate hierarchically, where the VDM is conditioned on the output from the flows (i.e., the halo and central galaxy properties).
In essence, the \nn framework takes in the simulation parameters \vthetasim as inputs and outputs properties of the halo, central galaxy, and satellites within the halo $\{\vthetahalo, \vthetacent, \vthetasat\}$. 
The target distribution of \nn, $p(\vthetasat, \vthetacent, \vthetahalo | \vthetasim)$, can be written as:
\begin{equation}
    p(\vthetasat | \vthetahalo, \vthetacent, \vthetasim)p(\vthetahalo, \vthetacent | \vthetasim).
\end{equation}
We note that the flows and diffusion models are trained independently, with more details outlined in \S\ref{section:training}.
A visual representation of \nn is shown in Figure~\ref{fig:flow_chart}.

We note that this hierarchical approach is a deliberate design choice.
One could, for example, model the joint likelihood of central and satellite galaxies, i.e., $p(\vthetasat, \vthetacent | \vthetahalo, \vthetasim)$, using the VDM.
We choose to separate the central and satellite galaxies for the following reasons:
\begin{itemize}
    \item It is common to study the relationships between central galaxies and their satellites. 
    By modeling the conditional likelihood of satellite galaxies given the central galaxy (instead of the joint likelihood of satellites and the central galaxy), we can more easily adjust the properties of the central galaxies to observe how these changes influence the satellites.
 
    \item This is especially important for the ``isolated Milky Way'' sample in the TNG-WDM suite of simulations used in this paper. 
    The central galaxy dominates the mass within the halo and is significantly more massive than its satellites, typically by a few orders of magnitude.
    We thus expect it to play a significant role in the dynamics of satellites through processes such as tidal stripping, dynamical friction, and mergers, which can alter the mass distribution within the halo and affect the formation and evolution of satellites~\citep{2005Natur.435..629S, 2008gady.book.....B, 2010gfe..book.....M}.
\end{itemize}

In addition, the normalizing flows could be further separated into two distinct components, working hierarchically: one flow that models halos, i.e., $p(\vthetahalo | \vthetasim)$, and another that models central galaxies given their halos, i.e., $p(\vthetacent | \vthetahalo, \vthetasim)$. 
As with satellite and central galaxies, this hierarchical structure could potentially allow for a more modular and flexible approach to modeling the different components within the halo.
However, we note that this separation is not critical, as the VDM ultimately takes both $\vthetacent$ and $\vthetahalo$ as inputs. 
Our emphasis in this paper is on the VDM approach, which we believe is the core component of our HOD modeling framework.
We will further discuss applications of the \nn framework, along with potential modifications to the models and their features, in \S\ref{section:discussion}.

Lastly, we note that traditional HOD applications typically assign galaxies onto \textit{pre-existing} DM halos from N-body simulations, making the modeling $p(\vthetahalo | \vthetasim)$ using the flows unnecessary. 
However, given our limited number of simulations, this modeling approach facilitates more robust testing against the simulations, as it allows to generate new halos by drawing from the prior $p(\vthetasim)$.
We briefly discuss this in \S\ref{section:discussion}.

\subsection{Dataset and Features}
\label{section:dataset}

For each simulation, we extract the high-resolution Milky Way-mass halo together with its subhalos from the group catalogs as follows. 
First, we identify halos with a total mass within $4.84 \times 10^{11} \, \modoth < M_\mathrm{halo} < 2.07 \times 10^{12} \, \modoth$.
We require that the total mass of the low-resolution DM particles does not exceed $2\%$ of the total mass within the virial radius in all of our halos. 
From the 1,024 simulations in the TNG-WDM suite, 1,018 halos satisfy the contamination criteria.

As mentioned previously, \nn consists of two components.
Given a set of simulation parameters \vthetasim, we first predict some properties of the halo \vthetahalo and the central galaxy \vthetacent.
Then, given the halo, central galaxy, and simulation parameters, we predict some properties of all satellite galaxies within the halo \vthetasat.
Below, we provide a detailed description of the set of properties that go into $\{\vthetasim, \vthetahalo, \vthetacent, \vthetasat\}$.
Table~\ref{tab:features} provides a summary of the halo and galaxy features and the corresponding \textsc{FoF/subfind} field in the public data release.
Descriptions of the simulation parameters can be found in Table~\ref{tab:prior}.

\subsubsection{Simulation parameters}
The simulation parameters \vthetasim are the input features of the entire \nn framework, i.e. both the normalizing flows and the VDM.
Here, we let \vthetasim be the WDM mass and three astrophysical parameters that control the strength of baryonic feedback:
\begin{equation}
    \vthetasim = \{\mwdm, \sno, \snt, \agn \}.
\end{equation}
In practice, we use the inverse of the WDM mass and the logarithm of the feedback parameters, since they are sampled uniformly (according to a Sobol sequence) in this space (see Table~\ref{tab:prior}.

We note that the properties of halos and central galaxies can depend on factors beyond these four parameters, such as environmental (e.g., local density field) and assembly information (e.g., star formation history, formation timescales)~\cite[e.g.,][]{1999MNRAS.303..685N, 2002MNRAS.331...98V, 2006ApJ...652...71W, 2015MNRAS.450.1521C, Hadzhiyska2021, 10.1093/mnras/stac2297}.
As discussed in \S\ref{section:sim_ic}, each halo in our training simulations represents a unique realization of a Milky Way-mass halo, with the only environmental constraint being the isolation criterion.
We expect that including dependencies on environmental and assembly factors could further improve the model's accuracy, which we plan to explore in future work.

\subsubsection{Halos and frame of reference}
The normalizing flows take the simulation parameters \vthetasim as inputs and outputs the properties of the halo 
\vthetahalo and central galaxy \vthetacent. 
For the halo properties, \vthetahalo, we consider the halo virial mass \mhalo, total stellar mass \mstar, and the number of satellite subhalos \nsat, i.e. 
\begin{equation}
        \vthetahalo =  \{\mhalo, \mstar, \nsat\}.
\end{equation}
The virial mass \mhalo is estimated as the total mass enclosed within $R_{200}$, the radius where the average density of the halo is 200 times the critical density of the Universe.  
For each halo, we use a reference system where its center and peculiar velocity within the cosmological box are both at $(0, 0, 0)$.
The halo center is the position of the particle with the minimum gravitational potential energy, while the velocity is computed as the sum of the mass-weighted velocities of all particles in the \textsc{FoF} group.

Normalizing flows are designed to model distributions of continuous variables and thus are ill-suited to predict discrete variables such as the number of satellites \nsat.
Thus, in practice, we instead train on and predict $\log \nsat$.
During generation, we compute \nsat from the predicted $\log \nsat$ and then round to the nearest integer.
We discuss the limitation of this approach in \S\ref{section:discussion}.

\subsubsection{Central galaxies}
For properties of the central galaxy, \vthetacent, we are interested in modeling the total mass, stellar mass, DM concentration, and the position and velocity offsets relative to the halo.
For the position offset, we find that central galaxies in our simulations are at $(0, 0, 0)$ using the halo reference frame described above.
Thus, we do not include the position offset into our features.
For the velocity offset, we find a small offset of order 5-10 km/s between the velocities of each halo and its central galaxy. 
Assuming this velocity offset is isotropic\footnote{This is reasonable because the gravitational forces acting upon the central galaxy from various directions tend to average out due to the symmetric distribution of matter within the halo, resulting in an overall isotropic velocity distribution.}, we only need to model its magnitude, i.e. $|\mathbf{v}_\mathrm{cent}|$.

In addition to the total mass $M_\mathrm{cent}$, stellar mass $M_\mathrm{cent, \star}$, and the velocity offset $|\mathbf{v}_\mathrm{cent}|$, we are interested in modeling structural parameters, such as the DM concentration, which has been found to strongly correlate with the halo formation time and environment in many assembly bias studies~\cite[e.g.,][]{1999MNRAS.303..685N, 2002MNRAS.331...98V, 2006ApJ...652...71W, 2015MNRAS.450.1521C}.
Because the DM concentration is not available in \textsc{subfind}, we instead use the quantity:
\begin{equation}
    \vmaxtildei \equiv \frac{V_{\mathrm{max}, i}}{H_0 r_{\mathrm{max}, i}},
\end{equation} 
where $H_0$ is Hubble constant, $V_{\mathrm{max}, i}$ is the maximum circular velocity of the galaxy and $r_{\mathrm{max}, i}$ is the radius at this velocity.
This follows from~\cite{2019MNRAS.490.5693B}, which demonstrates that $\tilde{V}_\mathrm{max}$ is a good proxy for the DM concentration.
We also highlight that the relation between \vmaxtilde and the DM concentration $c$ of a Navarro-Frenk-White~\cite[NFW,][]{1997ApJ...490..493N} is one-to-one (see Equation 9 of  \citealt{2008MNRAS.391.1685S}), so both \vmaxtilde and $c$ can be used interchangeably as training features. 
Our post-processing catalog will include both the concentration proxy \vmaxtilde and the concentration $c$. 

To summarize, the properties of central galaxies used as training features in this work are:  
\begin{equation}
    \vthetacent = \{M_\mathrm{cent}, M_\mathrm{cent, \star}, |\mathbf{v}_\mathrm{cent}|, \tilde{V}_\mathrm{max, cent} \}.
\end{equation}
These features, along with the simulation parameters \vthetasim and the halo properties \vthetahalo, are used as the context of the VDM for generating satellite galaxies.
As before, we emphasize that environmental information and assembly history are not explicitly included as training features. 
Although the DM concentration proxy \vmaxtilde is likely correlated with these factors, it would be beneficial to explicitly include environmental and assembly information in future work.

\subsubsection{Satellite galaxies}
For each satellite galaxy, our model predicts the 3-dimensional positions $\mathbf{x}_i$ and velocities $\mathbf{v}_i$ relative to the halo, the total mass \msubhaloi, the stellar mass \msubhalostari, and the concentration proxy \vmaxtildei.
The features of each subhalo, denoted as $\vtheta_i$, are thus:
\begin{equation}
    \vtheta_i = \{\mathbf{x}_i, \mathbf{v}_i, \msubhaloi, \msubhalostari, \vmaxtildei \},
\end{equation}
forming a 9-dimensional output vector for each galaxy. 
Additionally, it is interesting to include other internal properties of the satellite galaxies (e.g., spin, halo shape, galaxy morphology, etc.) into the framework. 
We leave this for future work.

Our framework outputs a vector $\vthetasat = \{\vtheta_i | i=1,..., \nsat\}$ of shape $\nsat\times9$ containing the properties of \nsat satellites.
The number of satellites is a function of each of the properties of the context $\vthetasim$ (e.g., a larger value of \mwdm will result in more satellites per halo). 
However, we note that due to differences in formation history and environment, \nsat can vary between halos for the same \vthetasim.
This makes the use of permutation equivariant architectures such as Transformers advantageous, as will be discussed in more detail in \S\ref{section:vdm}.

\begin{table*}
    \centering
    \begin{tabular}{llll}
        \hline
        & Parameters & Description & Relevant FoF/SubFind Field \\ 
        \hline
        Halo Properties & \mhalo & Halo virial mass & \texttt{Group\_M\_Crit200} \\ 
        \vthetahalo & \mstar & Halo stellar mass & \texttt{GroupMassType PartType4} \\
        & \nsat & Number of satellites &  \\
        \hline
        Central Galaxy  & $|\mathbf{v}_\mathrm{cent}|$ & Velocity offset with respect to the halo & \texttt{SubhaloVel} \\
        Properties \vthetacent & $M_\mathrm{cent}$ & Total mass of central galaxy & \texttt{SubhaloMass} \\
        & $M_\mathrm{cent, \star}$ & Stellar mass of central galaxy & \texttt{SubhaloMassType PartType4} \\
        & $\Tilde{V}_\mathrm{max, cent}$ & Proxy for DM concentration of the central galaxy  &   \\
        \hline
        Satellite Galaxy & $\mathbf{x}_i$ & 3-dimensional position vector of satellite $i$ &\texttt{SubhaloPos} \\
        Properties \vthetasat & $\mathbf{v}_i$ & 3-dimensional velocity vector of satellite $i$ &\texttt{SubhaloVel} \\
        & \msubhaloi & Total mass of satellite $i$ & \texttt{SubhaloMass} \\
        & \msubhalostari & Stellar mass of satellite $i$ & \texttt{SubhaloMassType PartType4} \\
        & $\Tilde{V}_\mathrm{max, i}$ & Proxy for DM concentration of satellite $i$  & \\
        \hline
    \end{tabular}
    \caption{List of parameters used in this work and their corresponding \textsc{FoF/subfind} fields, if available, in the public data release. 
    Note that the properties of central galaxies are extracted from the same \textsc{FoF/subfind} fields as satellite galaxies, but with different indices. 
    The indices of central galaxies are found at \textsc{GroupFirstSub} in the group catalog.}
    \label{tab:features}
\end{table*}

\subsection{Machine learning framework}
\label{section:vdm}

In this section, we describe the normalizing flows and VDM used in \nn, including their motivations and neural network architectures.
The mathematical formulations of normalizing flows and diffusion models are comprehensive and thoroughly documented in the machine learning literature. 
Thus, we omit their detailed derivations from this section. 
We instead provide the mathematical formalism of both models, including their optimization objectives, in Appendix~\ref{app:flows} for normalizing flows and Appendix~\ref{app:vdm} for diffusion models. 

\subsubsection{Modeling halo and central galaxies}
To model the properties of halos and central galaxies, we employ normalizing flows~\citep{2015arXiv150505770J, 10.5555/3294771.3294994, 2019arXiv191202762P} that are conditioned on simulation parameters $\vthetasim$.
Normalizing flows employ non-linear, invertible transformations (with tractable Jacobians and its inverse) from a standard, base Gaussian distribution to a complex target distribution, allowing for efficient density estimation and sampling.
They have been employed in various astrophysical applications including generative modeling~\cite[e.g.,][]{Lovell_2023, 2023arXiv230805145N, Kwon_2024} and simulation-based inference~\cite[see][for a review]{2020PNAS..11730055C}.

In our framework, our flow model consists of 8 neural spline flows \citep{2019arXiv190604032D} layers, each using a rational quadratic spline transformation with 8 bins (9 knots). 
To read in the conditioning context, each flow also has a multi-layer perceptron (MLP) with 2 fully-connected layers of hidden size 16 and \texttt{ReLU} activation function. 
Each fully-connected layer is followed by a batch normalization layer~\citep{2015arXiv150203167I} and a dropout layer~\citep{JMLR:v15:srivastava14a} with a rate of 0.2.
We find that, due to the limited number of simulations available in our application, batch normalization and dropout are especially important to prevent overfitting in the flows.
During training, we also employ early stopping to prevent overfitting (see \S\ref{section:training}). 

\subsubsection{Modeling satellite galaxies}
To generate satellites within a halo, we adopt an approach presented in \citetalias{cosmovdm}. 
We represent each satellite population as a point cloud---a set of points in 3-dimensional space with attributes such as velocities, masses, and concentrations.
Point clouds eliminate the need for binning and voxelization, which are not viable/effective for HOD applications due to the sparse spatial distribution of satellites. 
More importantly, this allows us to retain the most information from the data and resolve arbitrary small scales limited only by the spatial resolution of the simulations.

Similarly, we also employ a Variational Diffusion Model~\cite[VDM;][]{2021arXiv210605931V, 2021arXiv210109258S, 2021arXiv210700630K}.
VDMs are a class of generative models that progressively learn to generate complex data distributions through a process of gradually denoising samples (referred to as the ``reverse diffusion''), starting from a random noise distribution.
They offer a few advantages over other generative models.
VDMs are more expressive than variational autoencoders,~\cite[VAEs;][]{2013arXiv1312.6114K}, allowing for more flexible network architecture than normalizing flows.
Unlike generative adversarial networks,~\cite[GANs;][]{2014arXiv1406.2661G}, they are generally much easier to train and less prone to experiencing mode collapse~\citep{2016arXiv160603498S}.
Additionally, they facilitate the tractable evaluation of likelihood, which allows them to be used for inference and anomaly detection.

The VDM consists of a \textit{noise schedule}, $\gamma(t)$, which determines how much Gaussian noise is added to the data $\mathbf{y}$ at each step $t$ of the forward diffusion process, and a \textit{noise prediction model}, $\vect{\hat{\epsilon}}(\zzt{t}, t)$, which estimates the added noise given the time step $t$ and the noisy data \zzt{t}~\citep{2020arXiv200611239H, 2020arXiv201113456S, 2021arXiv210700630K}.
In our setup, we use a linear noise schedule with trainable parameters $\eta_\mathrm{min}$ and $\eta_\mathrm{max}$ that determine the minimum and maximum amount of noise added, respectively.
Further details on the diffusion process, including the noise schedule, and forward and reverse diffusion process, can be found in Appendix~\ref{app:vdm}.

For the noise prediction model, we use a Transformer-based architecture.
The Transformer, first introduced in~\cite{vaswani2023attention}, leverages the attention mechanism to efficiently process data.
Attention allows Transformer-based models to dynamically assess and prioritizes the relevance of each part of the input, making them powerful models for many applications ranging from large language modeling~\citep{radford2019language, 2018arXiv181004805D} to more recent astrophysics applications \citep[see e.g.][]{2023JCAP...11..075H, 2023PASJ...75.1311H, 2023arXiv231003024L, cosmovdm, 2024MNRAS.527.1494L}.

We implement a Transformer with additional MLPs for processing conditioning features (i.e., $\{\vthetasim, \vthetahalo, \vthetacent\}$), the diffusion noise time step $t$, and the input data  (i.e., the noisy version of \vthetasat at $t$).
During forward pass, the time step $t$ is first passed through a sinusoidal positional encoding layer similar to that used in~\cite{vaswani2023attention}.
The encoded time step, the conditioning features, and the input data are then independently projected onto a 128-dimensional space by three 1-layer MLPs.
The projected conditioning features and time step are further concatenated and passed through a 2-layer MLP.
This output is then added to the projected input data before being passed into the Transformer. 
The Transformer itself consists of 6 blocks, each equipped with a Multi-head Attention layer with 128 hidden channels and 4 attention heads, followed by a 2-layer MLP with 256 hidden channels and a \texttt{GELU} activation function~\citep{2016arXiv160608415H}. 
The output is then mapped back into the same dimension of the input data using a 1-layer MLP.  

Because each satellite population is modeled as a set of points with no inherent ordering, we do not include any position encoding or casual masking layer, making the model permutation-equivariant.
It is also important to highlight that our neural network architecture does not explicitly incorporate geometrical symmetries, such as rotational and translational invariance. 
Instead, we apply data augmentation techniques during the training process, as detailed in \S\ref{section:training}.

Lastly, it is worth noting that graph-based neural networks are also permutation-equivariant and capable of efficiently modeling point cloud cosmological data, as have been shown in recent studies~\cite[e.g.,][]{2022ApJ...937..115V, 2023PhRvD.107d3015N, 2023ApJ...952...69D, 2024MNRAS.527..739R, dreams}.
\citetalias{cosmovdm} also demonstrates that using a graph convolutional neural network and $k$-nearest neighbors graphs can yield comparable results to Transformers.
An advantage of using graph neural networks is that the computational cost scales more slowly with the number of point cloud data (i.e. the number of satellites) increase, as self-attention scales quadratically $\mathcal{O}(\nsat^2)$.
However, in our application, where the number of points is of order $\nsat \sim \mathcal{O}(100)$, the performance difference becomes negligible. 
In addition, constructing graphs from point clouds introduces additional hyperparameters, e.g., the number of neighbors $k$. 
For these reasons, we thus choose a Transformer-based architecture for our noise prediction model.

\subsection{Training and Optimization}
\label{section:training}

\begin{figure*}
    \includegraphics[width=0.95\linewidth]{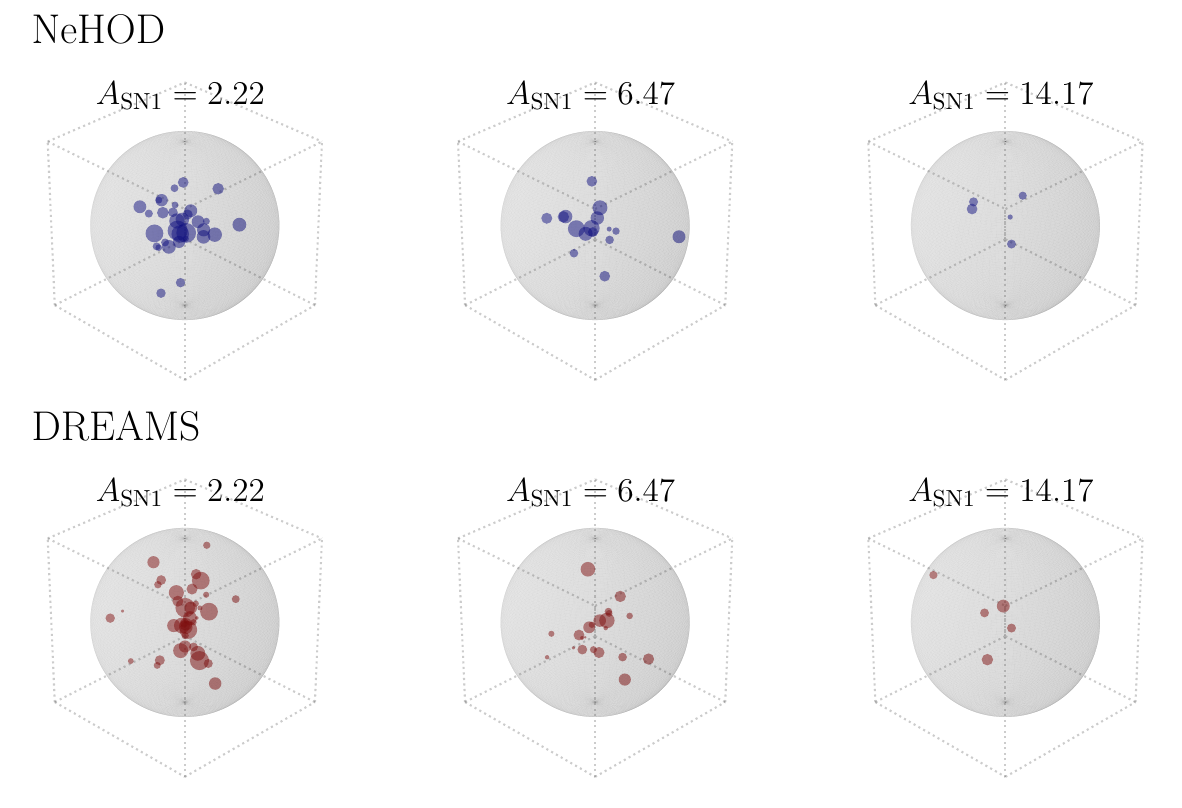}
    \caption{Example satellite galaxies generated by \nn. 
    The bottom row displays the satellite galaxies of three halos from the DREAMS simulations, with each column corresponding to a different value of \sno. 
    The top row shows corresponding realizations of satellite galaxies for the same halos, as generated by \nn. 
    The marker size scales logarithmically with the stellar mass of each satellite. 
    For visual clarity, central galaxies, located at the center of the shaded sphere, are omitted. In each panel, the box size and the diameter of the shaded sphere are set to $600 \, \mathrm{kpc} \, h^{-1}$.
    }
    \label{fig:example_samples}
\end{figure*}

We split the dataset into 814 training and 204 validation simulations, following an 80-20 split. 
The flows and the VDM are trained independently by performing gradient descent using the AdamW optimizer~\citep{adamw2019, kingma2014adam}, with a peak learning rate of  $0.0005$ and a weight decay coefficient of $0.01$.
Details on the training objectives can be found in Appendix~\ref{app:flows} for the flows and Appendix~\ref{app:vdm} for the VDM. 
In addition, we use a cosine decay learning rate scheduler~\citep{2016arXiv160803983L}.
Training parameters, including the linear warm-up and cosine decay steps of the scheduler and the training batch size, are $\{5000, 10000, 128\}$ and $\{100, 1000, 64\}$ for the VDM and the flows, respectively.
The VDM and flows converge after about 50,000 (400) and 10,000 (150) iterations (epochs), which take approximately 30 and 20 minutes for each model on an NVIDIA Tesla V100 GPU.
We implement early stopping with the patience of 1000 iterations for both models to prevent overfitting and select the checkpoints with the lowest validation loss.

As mentioned in \S\ref{section:vdm}, we apply data augmentation to help the model learn the relevant geometrical symmetries, such as translational and rotational invariance.
Data augmentation also mitigates the limitations of our relatively small training dataset by effectively increasing the diversity of the training examples. 
We consider only rotational invariance because the coordinates are always in the frame of the halo by construction (as discussed in \S\ref{section:dataset}).

The data augmentation is applied \textit{during training}.
In other words, a random 3-dimensional rotational matrix is applied to the coordinates of the satellites on the fly before being passed through the VDM. 
\textit{Thus, it is unlikely that the model ever encounters the same sample twice, which improves generalization to unseen data.}
The same approach is applied during validation, ensuring that the model is consistently tested on varied data configurations.
We do not consider planar symmetry (e.g., the disk plane of the central galaxy) in this work. 
Depending on specific applications, future work may incorporate planar symmetry into the data augmentation process. 

\section{Results}
\label{section:result}

We present our results by generating new samples using \nn and validating them by comparing key summary statistics of the generated samples with those from the DREAMS simulations.
To generate the test dataset, we randomly sample the WDM mass and astrophysical parameters from the prior distributions in Table~\ref{tab:prior}.
We first generate the halos and central galaxies using the conditional flows, and then use the VDM to generate satellites given the generated halos and central galaxies.
In total, we generate 100,000 samples, which take approximately 50 minutes on a GPU using 1,000 diffusion steps for each sample.
Note that the generated sample size is 100 times larger than the sample size of the training simulations (about 1000 samples), which will help identify potential discrepancies and biases between the two datasets more effectively.

Figure~\ref{fig:example_samples} presents an example output of \nn. 
In the bottom row, we display the satellite galaxies of three halos randomly selected from the DREAMS simulations, with increasing \sno parameter from left to right. 
For each halo, we generate a corresponding realization of its satellite galaxies using the same halo properties (including the number of satellites), central galaxy properties, and simulation parameters, and plot them in the top panel of the corresponding column. 
For visual clarity, we do not show the central galaxies, which are at the center of the shaded sphere.
Figure~\ref{fig:example_samples} effectively highlights that the point-cloud output of \nn, with more detailed validation tests presented in the rest of the section.

\subsection{Properties of halos and central galaxies}
\label{section:res_halo_prop}

\begin{figure*}
    \centering
    \includegraphics[width=\linewidth]{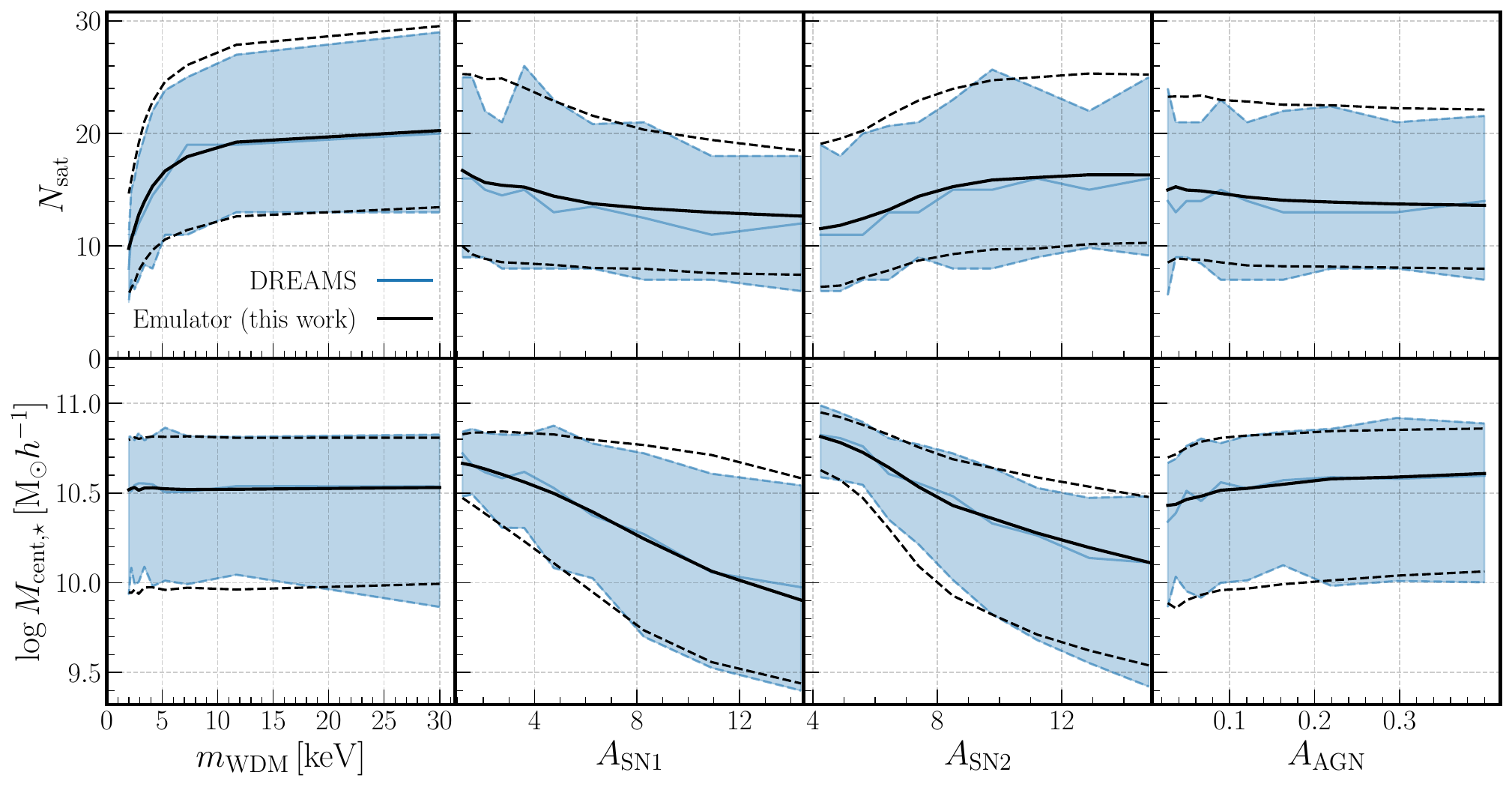}
    \caption{
    The properties of halos and central galaxies as a function of the simulation parameters \simparam (left to right).
    The top and bottom panels show distributions of the number of satellites \nsat and the stellar mass of the central galaxies \mcentstar. 
    The median (solid lines), 16th, and 84th percentiles (dashed lines) of each distribution are shown. 
    The black lines denote the samples generated by conditional flows in this work, while the blue lines and shaded regions denote the simulations.}
    \label{fig:flows}
\end{figure*}

\begin{figure*}
    \centering
    \includegraphics[width=0.8\linewidth]{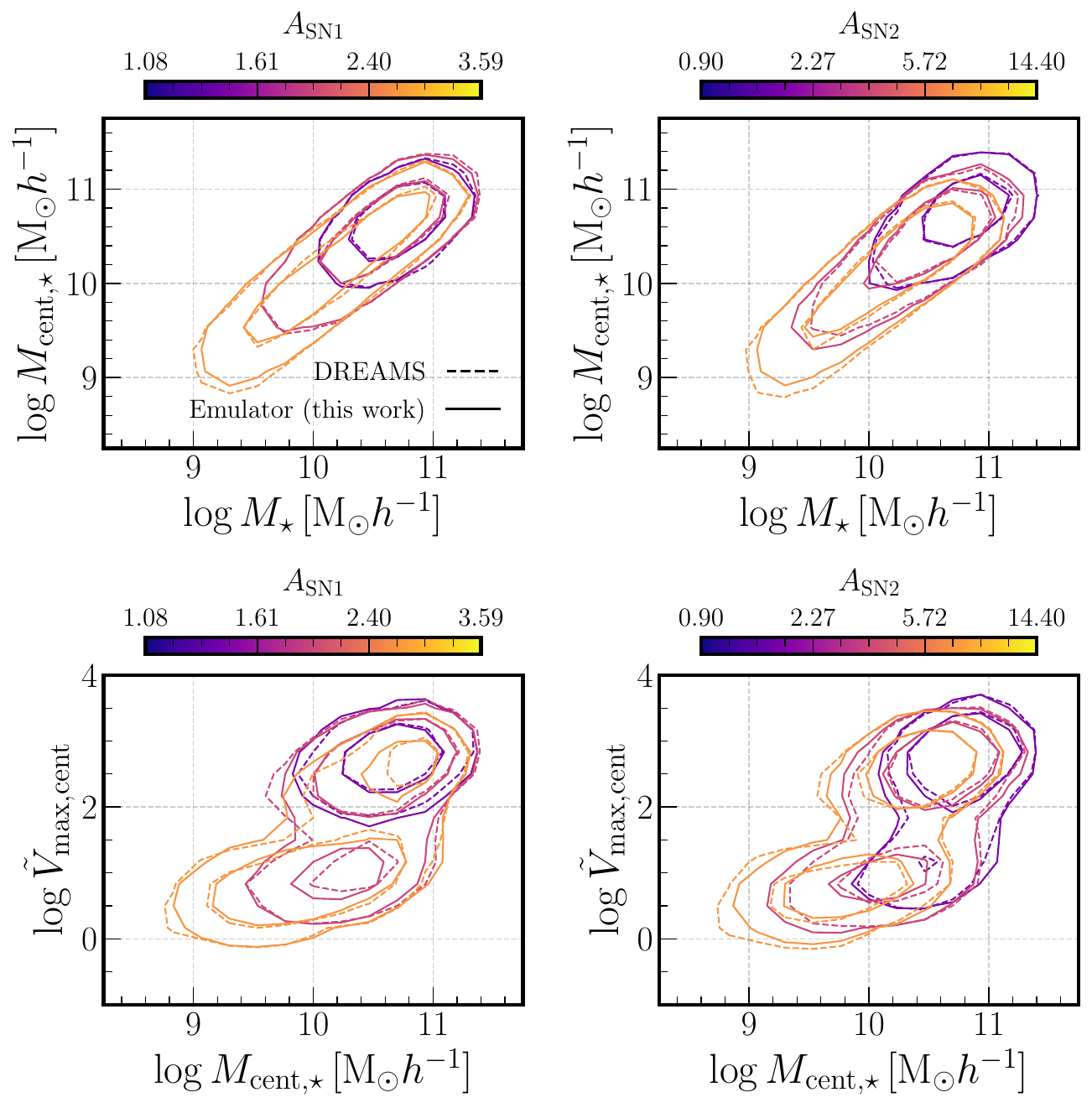}
    \caption{
    \textit{Top:} distributions of the stellar masses of the halos \mstar and central galaxies \mcentstar. 
    \textit{Bottom:} distributions of the central galaxy stellar mass \mcentstar and the concentration proxy \vmaxtilde.
    In both cases, the 68\% and 95\% contours are shown for 3 bins of \sno (left) and \snt (right), with each bin denoted by a different color.
    The solid and dashed lines denote the \nn samples and the simulations, respectively.
    }
    \label{fig:flows_nd}
\end{figure*}

We evaluate how well the conditional flows can recover the properties of the halos and central galaxies. 
We divide the samples into 10 bins of the simulation parameters \simparam.
For each property of halos and central galaxies (i.e., \vthetahalo, \vthetacent), we calculate the median, 16th, and 84th percentiles of its distribution within each bin.
The bins are divided inverse-uniformly for the WDM mass and log-uniformly for the astrophysical parameters (see Table~\ref{tab:prior}) and thus have the same sample size. 
In particular, each bin contains about 1,000 and 100 halos for the \nn dataset and the simulations, respectively.

Figure~\ref{fig:flows} shows variations of the number of satellites \nsat (top) and central galaxy stellar mass \mcentstar (bottom) with respect to the simulation parameters.
We show the full set of output features in Appendix~\ref{app:res_flows}.
The halo virial mass \mhalo and central galaxy total mass \mcent does not vary significantly with these parameters, which is likely because the halos are selected based on the Milky Way-mass criterion (see \S\ref{section:dataset}).
Thus, we do not show the results for \mhalo and \mcent, and briefly note that the distributions of these properties predicted by \nn match well with those from the simulations.
Additionally, since the stellar masses of the halo and central galaxy are strongly correlated, we exclude the halo stellar mass \mstar here and instead show their 2-dimensional contours in Figure~\ref{fig:flows_nd}.
Lastly, the concentration proxy \vmaxtilde follows a bimodal distribution that strongly correlates with \mcentstar, rendering the above statistics less informative. 
Thus, as with the halo stellar mass, we show the 2-dimensional contours of \vmaxtilde and \mcentstar in Figure~\ref{fig:flows_nd}.

In Figure~\ref{fig:flows}, we observe strong variations in the number of satellites \nsat and the central galaxy stellar mass \mcentstar with respect to the simulation parameters, except for \agn. 
The number of satellites \nsat is sensitive to both the WDM mass and the SN feedback parameters, while \mcentstar is primarily influenced by the latter. 
Overall, there is good agreement between the distributions predicted by \nn and those from the simulations. 
However, the flows tend to slightly overestimate \nsat by 1-2 satellites, though this discrepancy is typically much smaller than the scatter in the \nsat distribution due to halo-to-halo variations.
On the other hand, the distributions of \mcentstar show good agreement for \mwdm, \sno, and \agn, with the largest discrepancy occurring in the 16th percentile for \snt.
We note that we do not expect the flows to agree perfectly with the simulations, as the number of training samples is relatively small ($\approx 1000$ simulations).
This is especially true towards the edges of the target distributions, where the models require more training samples to learn (as compared to the median).
We expect that with more simulated data, the accuracy of the distributions will further improve.

Figure~\ref{fig:flows_nd} shows the 2-dimensional distributions of \mstar and \mcentstar (top), and \mcentstar and \vmaxtilde (bottom).
The \nn samples and simulations are each divided into 3 bins of the simulation parameters (denoted by different colors), each containing approximately 33,000 and 330 samples respectively.
The left and right panels show the 68\% and 95\% contours of the distributions for the \sno and \snt bins, respectively.
We do not show bins of \mwdm and \agn as we do not find strong variations of \mcentstar, \mstar, and \vmaxtilde with respect to these parameters.
Note that due to the differences in the sample size, the simulation contours are noticeably noisier than the \nn contours. 

As briefly mentioned previously, there are strong correlations between \mstar and \mcentstar, and between \mcentstar and \vmaxtilde.
The distribution of the concentration proxy \vmaxtilde is bimodal, which could be attributed to the bimodality in the formation history of DM halos~\citep[e.g.,]{2018ApJ...857..127S, 2021OJAp....4E...7H}.
Since the concentration is highly correlated with formation history~\cite[e.g.,][]{1999MNRAS.303..685N, 2002MNRAS.331...98V, 2006ApJ...652...71W, 2015MNRAS.450.1521C}, the observed bimodality in \vmaxtilde can arise from differences in the formation histories of these halos.
A more detailed investigation of this bimodality is beyond the scope of this work, so we leave this for future work.

From Figure~\ref{fig:flows_nd}, we observe small differences in the case of the central galaxy stellar mass \mcentstar and concentration proxy \vmaxtilde, e.g. the 68\% contours in the \sno (lower left) panel. 
As previously noted, we do not expect the distributions to match perfectly, due to the limited number of training samples.
Bimodal distributions are also much more challenging to fully capture, as both peaks need to be well-sampled to accurately represent the target distribution. 
Overall, however, the 68\% and 95\% contours predicted by \nn generally align well with the simulations.

Lastly, we note that \S 3 of \citealt{dreams} presented a machine learning-based emulator for the WDM satellite count. 
Given the four simulation parameters, the authors assume a Gaussian distribution of satellite count and employ MLPs to predict the mean and standard deviation. 
This MLP approach also allows both efficient sampling and likelihood evaluation and achieves somewhat similar results to the conditional flows.
However, we opt to use the flows in our framework as they require fewer assumptions on the target distributions and thus can model more complex distributions. 
As seen in Figure~\ref{fig:flows_nd}, properties such as the concentration proxy \vmaxtilde display non-Gaussian features, i.e. bimodal distributions that strongly correlate with the stellar mass of the central subhalos.
In such cases, we expect the flows to outperform the MLP approach employed in \citealt{dreams}.

\subsection{Properties of satellite galaxies}
\label{section:res_sat_prop}

We now evaluate the properties of the satellite populations generated by the VDM, by comparing their key summary statistics with those derived from the simulations.
We divide each parameter into three bins such that each bin contains approximately the same number of samples. 
Due to the prior distributions in the simulation parameters (as shown in Table~\ref{tab:prior}), the bins are distributed uniformly in inverse WDM mass (i.e., \imwdm) and log-uniformly for the astrophysical parameters.
Each bin contains approximately 33,000 \nn-generated and 330 simulated satellite populations. 
This large generated sample size provides a more precise estimate of the distribution modeled by the VDM,

For each bin, we calculate and compare the satellite stellar mass functions (\S\ref{section:res_mf}), stellar-to-halo mass relations (\S\ref{section:res_shmr}), concentration-mass relations (\S\ref{section:res_cvir}), and position and velocity clustering (\S\ref{section:res_position}).
This provides validation tests of the VDM for modeling both intra-galaxy properties (i.e., properties within a single galaxy), such as the stellar-to-halo mass relations and concentration-mass relations, and inter-galaxy properties (i.e., between galaxies in a catalog), such as the satellite stellar mass functions and clustering. 

\subsubsection{Satellite stellar mass functions}
\label{section:res_mf}

\begin{figure*}
    \centering
    \includegraphics[width=\linewidth]{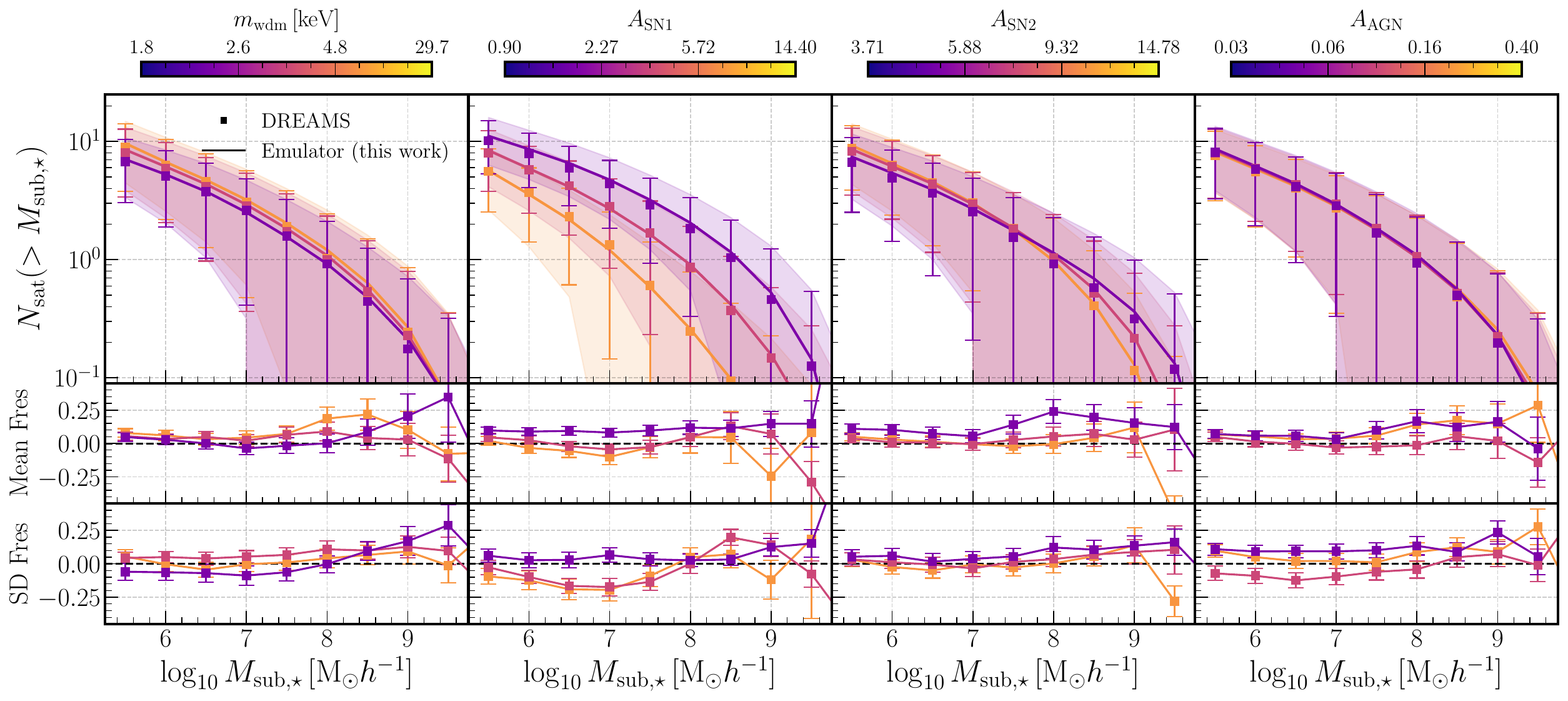}
    \caption{
    \textit{Top}: 
    The satellite stellar mass functions (SSMFs) generated by \nn and extracted from the simulations.
    The columns show the variations of the SSMFs over the WDM mass \mwdm and astrophysical parameters \astroparam.
    In each column, the color denotes the SSMFs of a bin of the corresponding parameter. 
    The average SSMFs, along with their standard deviations, are shown as solid lines and shaded regions for \nn and error bars for the simulations.  
    \textit{Middle}: 
    Fractional residuals of the average SSMFs, calculated as the difference between the \nn SSMFs and the simulation SSMFs divided by the simulation SSMFs. 
    The error bars represent the propagated errors of the fractional residuals, derived from the bootstrapped errors of the average SSMFs.
    \textit{Bottom}: 
    Fractional residuals of the standard deviation SSMFs, along with their errors, calculated using the same procedure as for the average SSMFs.
    }
    \label{fig:mf_stellar}
\end{figure*}

\begin{figure*}
    \centering
    \includegraphics[width=\linewidth]{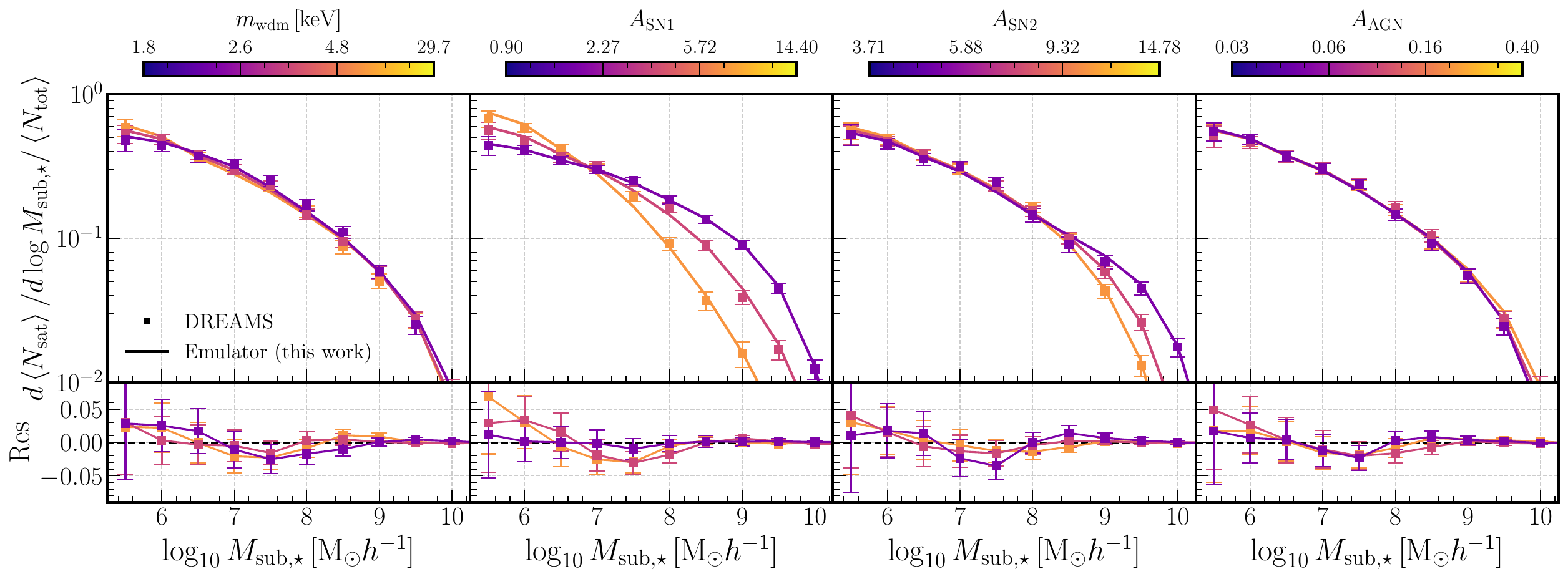}
    \caption{
    \textit{Top}: 
    The differential SSMFs $d \braket{\nsat} / d \log \msubhalostar$ normalized by the total number of satellites $\braket{N_\mathrm{tot}}$.
    The columns show the variations of the SSMFs over the WDM mass \mwdm and astrophysical parameters \astroparam.
    The \nn and simulation differential SSMFs are shown as solid lines and squares, respectively. 
    \textit{Bottom}: The residuals between \nn and simulation differential SSMFs.
    The error bars in both rows denote the errors of the simulations.     
    }
    \label{fig:mf_stellar_grad}
\end{figure*}

The satellite stellar mass function (SSMF) is among the key results of many Milky Way and dwarf galaxy surveys~\citep[e.g.,][]{2022ApJ...933...47C, 2023ApJ...956....6D, 2024arXiv240414498M} and has been used to place constraints on the mass of WDM~\cite[e.g.,][]{2014MNRAS.439..300L, 2018MNRAS.473.2060J, 2021ApJ...917....7N}.
Lower WDM particle masses result in fewer low-mass satellites in the SSMF due to free-streaming effects~\citep{2001ApJ...556...93B, 2014MNRAS.439..300L}. 
Baryonic feedback can also disrupt the formation of low-mass satellites and produce similar effects in the mass functions~\cite[e.g.,][]{2022NatAs...6..897S, 2024arXiv240311692B}. 
Thus, accurately modeling the SSMFs is important for differentiating the effects of DM from those of baryons and interpreting the results of these surveys.

We first calculate the SSMF of each halo by dividing its satellites into \msubhalo bins with equal widths of 0.5 dex and counting the cumulative number of satellites $\nsat(> \msubhalostar)$ in each bin. 
Then, for each bin of the WDM mass and astrophysical parameters, we calculate the average and standard deviation of the SSMF over all halos in the bin.
These statistics, along with their errors, are estimated using bootstrapping with 1,000 bootstrap samples, where each sample is generated by random sampling with replacement from the halos in each bin.
Note that due to the large number of generated halos ($\sim 33,000$ halos per bin), the errors of the generated SSMFs are a factor of $\sim 10$ smaller than that of the simulations and thus negligible.
In addition, for each parameter bin, we calculate the differential form of the SSMF, by taking the slope of the average SSMF $d\braket{\nsat}/d \log \msubhalostar$, which is another key summary statistic commonly reported by surveys~\citep{2022ApJ...933...47C, 2023ApJ...956....6D, 2024arXiv240414498M}.
It provides an important test of the VDM's ability to capture the shape of SSMF.
To further highlight the shape, we normalize by differential SSMF by the average total number of satellites in each parameter bin. 
We show additional results on the satellite halo mass functions in Appendix~\ref{app:mass_functions}.

Figure~\ref{fig:mf_stellar} shows the SSMFs for each bin of the WDM mass \mwdm and the three feedback parameters \astroparam (left to right column).
In each column, the top panel shows the average of SSMFs for the bins of the corresponding parameter, with each bin represented by a different color. 
The standard deviations of the SSMFs are shown as shaded regions and error bars for \nn and simulation samples, respectively. 
The middle and bottom panels show the fractional residuals of the averages and standard deviations of the SSMFs, respectively.
We show the fractional residuals instead of the residuals to highlight differences between \nn and the simulations for high \msubhalostar bins with low numbers of satellites. 
The error bars in the middle and bottom panels denote the errors of the fractional residuals, estimated using error propagation.

As expected, both the WDM mass and the feedback parameters (except for \agn) significantly influence the SSMFs.
There is a noticeable degeneracy between the effects of WDM mass and baryonic feedback and between the feedback parameters \sno and \snt. 
Additionally, the SSMFs are much more sensitive to \sno than to \mwdm, making it challenging to constrain \mwdm observationally with the SSHMFs due to the inherent uncertainty in \sno. 
This further highlights the need for precise modeling to disentangle these influences and place accurate constraints on WDM.

Overall, we see good agreements in both the average and standard deviation of the SSMFs between the simulations and generated samples, with the VDM accurately capturing the general trends across multiple bins of both the WDM mass \mwdm and the feedback parameters \astroparam. 
However, we observe a 10-15\% offset between the generated SSMFs and the simulations, which we attribute mainly to the differences in the total number of satellites \nsat.
As illustrated in Figure~\ref{fig:flows}, the normalizing flows tend to overestimate the total \nsat, leading to an overestimation at low \msubhalostar.
Additionally, the fractional residuals of the average SSMFs increase slightly at high \msubhalostar bins.
These bins have a lower number of satellites, and thus the average SSMFs fluctuate more noticeably.
The generated SSMFs remain within 1-2 error bars of the simulations, as the errors also increase at high \msubhalostar.
We observe similar trends with the standard deviations.

Figure~\ref{fig:mf_stellar_grad} shows the normalized differential SSMFs (top panels),
and the residuals between the generated and simulated SSMFs (bottom panels) for the same parameter bins.
As previously noted, the normalized differential SSMFs are estimated by dividing the slope of the average SSMFs shown in Figure~\ref{fig:mf_stellar} by the average total number of satellites, i.e. $d\braket{\nsat}/d \log \msubhalostar / \braket{N_\mathrm{tot}}$.
The error bars denote the errors in estimating normalized differential SSMFs of the simulations.
The errors of the \nn SSMFs are not shown, as they are negligible due to the large number of generated samples.

The differential SSMFs highlight the shape of the mass distributions and provide an important test of the VDM's ability to capture these distributions. 
Overall, the generated and simulated SSMFs align well, with the residuals mostly within 1 error bar.
We note, however, that we observe a similar offset at low \msubhalostar, due to the overall number of satellites being overestimated by the flows.
Additionally, unlike in Figure~\ref{fig:mf_stellar}, the errors are higher for lower values of \msubhalostar. 
This is because the errors scale with the number of satellites in the mass bins, which is higher for lower mass subhalos.

\subsubsection{Stellar-to-halo mass relation}
\label{section:res_shmr}

\begin{figure*}
    \centering
    \includegraphics[width=\linewidth]{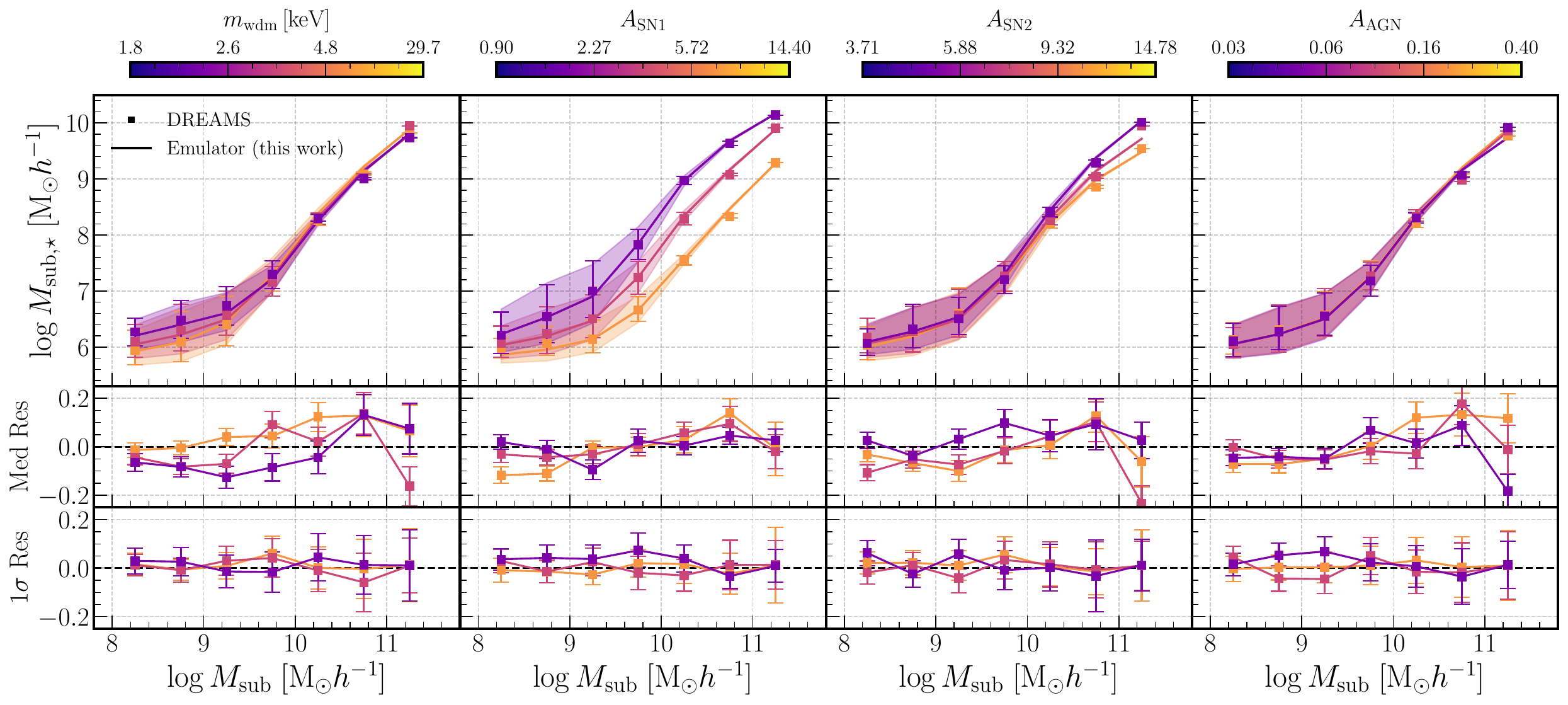}
    \caption{
    \textit{Top:}
    The stellar-to-halo mass relations (SHMR) of satellites generated by \nn and extracted from the DREAMS simulations.
    The columns show the variations of the SHMRs over the WDM mass \mwdm and astrophysical parameters \astroparam. 
    The average, 16th, and 84th percentiles of the SHMRs are shown as solid lines and shaded regions for \nn, and error bars for the simulations.
    \textit{Middle:}
    The residuals of the median SHMRs between \nn and the simulations, with the error bars denoting the errors of the simulations. 
    \textit{Bottom:}
    The residuals of the SHMR scatters between \nn and the simulations, with the error bars denoting the errors of the simulations. 
    }
    \label{fig:shmr}
\end{figure*}

The SSMFs provide an important test for capturing relationships between stellar masses of galaxies within a halo, i.e. the inter-galaxy properties. 
It is as important for the VDM to accurately model relationships between properties of a single galaxy, i.e. the intra-galaxy properties. 
To evaluate this aspect, we first compute the stellar-to-halo mass relation (SHMR).

The SHMR is a key summary statistic used in many galaxy formation studies~\citep[e.g.,][]{2017MNRAS.470..651R, 2018MNRAS.477.1822M, 2019MNRAS.488.3143B, 10.1093/mnras/stac2297}
The SHMR provides insights into the efficiency of star formation and the impact of feedback processes and thus is sensitive to baryonic prescriptions of the simulations~\citep[e.g.,][]{2000MNRAS.313..734E, 2014MNRAS.445..581H, 2015ARA&A..53...51S, 2019MNRAS.489.4233M}.
We note, however, that the SHMR has been employed mainly at the halo mass scales ($\gtrsim 10^{10} \, \modoth$) and not for satellite galaxies.
Nevertheless, it provides a good summary statistic for validating the relationships between stellar and halo mass predicted by \nn.

To compute the SHMR of the satellites within each halo, we divide the satellites into \msubhalo bins of 0.5 dex width. 
For each mass bin, we calculate the median, 16th, and 84th percentiles of the stellar mass \msubhalostar of the satellites. 
For each bin of simulation parameters \simparam, we then determine the average and standard error of these percentiles across all halos.
Additionally, we also estimate the 1-$\sigma$ confidence interval of the SHMRs in each parameter bin as the difference between the average 84th and 16 percentiles.

In Figure~\ref{fig:shmr}, the top rows show the average of the median, 16th, and 84th percentiles of the generated and simulated SHMRs in each parameter bin.
The solid lines and shaded regions represent the SHMRs of the \nn samples, while the error bars correspond to those of the simulations.
The middle and bottom panels present the residuals between the \nn and the simulations of the median and the scatter, respectively.
In both cases, the error bars denote the errors of the simulations.

As with the mass functions, we see that both the WDM mass and astrophysical parameters, except for \agn, affect the SHMRs, with the most dominant influence coming from \sno.
Interestingly, for low-mass satellites ($\msubhalo \lesssim 10^9 \, \mathrm{M_\odot} h^{-1}$), increasing the WDM mass \mwdm increases the subhalo stellar mass, though this is not a strong effect since all the lines are consistent to within $1\sigma$.
Similar effects have been observed in previous studies (e.g., \citealt{2017MNRAS.467.2019R, 2020MNRAS.498..702L}).
A potential explanation could be that WDM halos form later and thus have delayed star formation, resulting in a more extended period of gas availability for star formation. 
However, investigating these trends further is beyond the scope of this paper and is left for future work. 

The SHMRs of the generated satellites align well with those of the simulations. 
We see that \nn can capture the general trends in both the medians and 16-84th percentiles regions of the SHMRs.
Discrepancies from the median SHMRs of the simulations tend to increase slightly at high \msubhalo and typically about $0.1-0.2$ dex.
However, the uncertainties of the simulated median SHMRs also increase noticeably due to the limited number of high-mass satellites.
For example, there are approximately 30 simulated and 3000 generated samples that have satellites in the last mass bin of the SHMR.
Thus, even for this mass bin, the residuals of the median SHMRs between \nn and the simulations remain within 1-2 error bars of the simulations.

Next, we compare the average scatters of the SHMRs, estimated as the differences between the average 84th and 16th percentiles.
This is particularly important since second-order statistics, such as the SHMR scatters, contain a wealth of information about formation history of galaxies and cosmological~\citep[e.g.,][]{2018MNRAS.474.4089T, 2018MNRAS.478.2618F, 2020MNRAS.493.1361F, 2022ApJ...933...48F, 2020MNRAS.495..686A, Hadzhiyska2021, 10.1093/mnras/stac2297}.
From the bottom panels of Figure~\ref{fig:shmr}, we see a remarkable agreement between the \nn and the simulations.
This suggests that \nn can accurately capture not only the central trends but also the intrinsic scatter in the SHMR, which is crucial for understanding the underlying physics governing galaxy formation.

\subsubsection{Concentration-mass relation}
\label{section:res_cvir}

\begin{figure*}
    \centering
    \includegraphics[width=\linewidth]{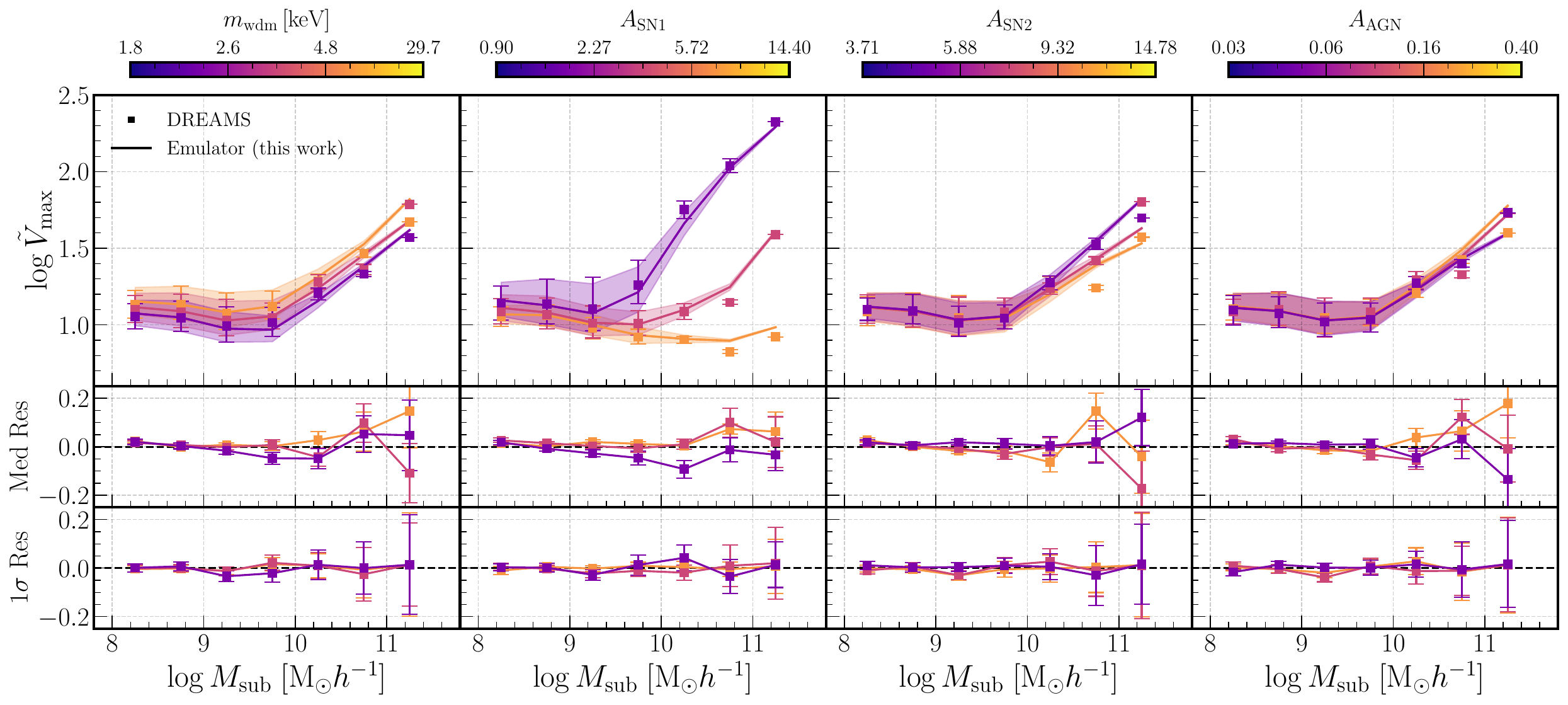}
    \caption{The concentration-mass relations of satellites generated by \nn and extracted from the DREAMS simulations.
    Panels are the same as in Figure~\ref{fig:shmr}. 
    }
    \label{fig:mcvir}
\end{figure*}

\begin{figure}
    \centering
    \includegraphics[width=\linewidth]{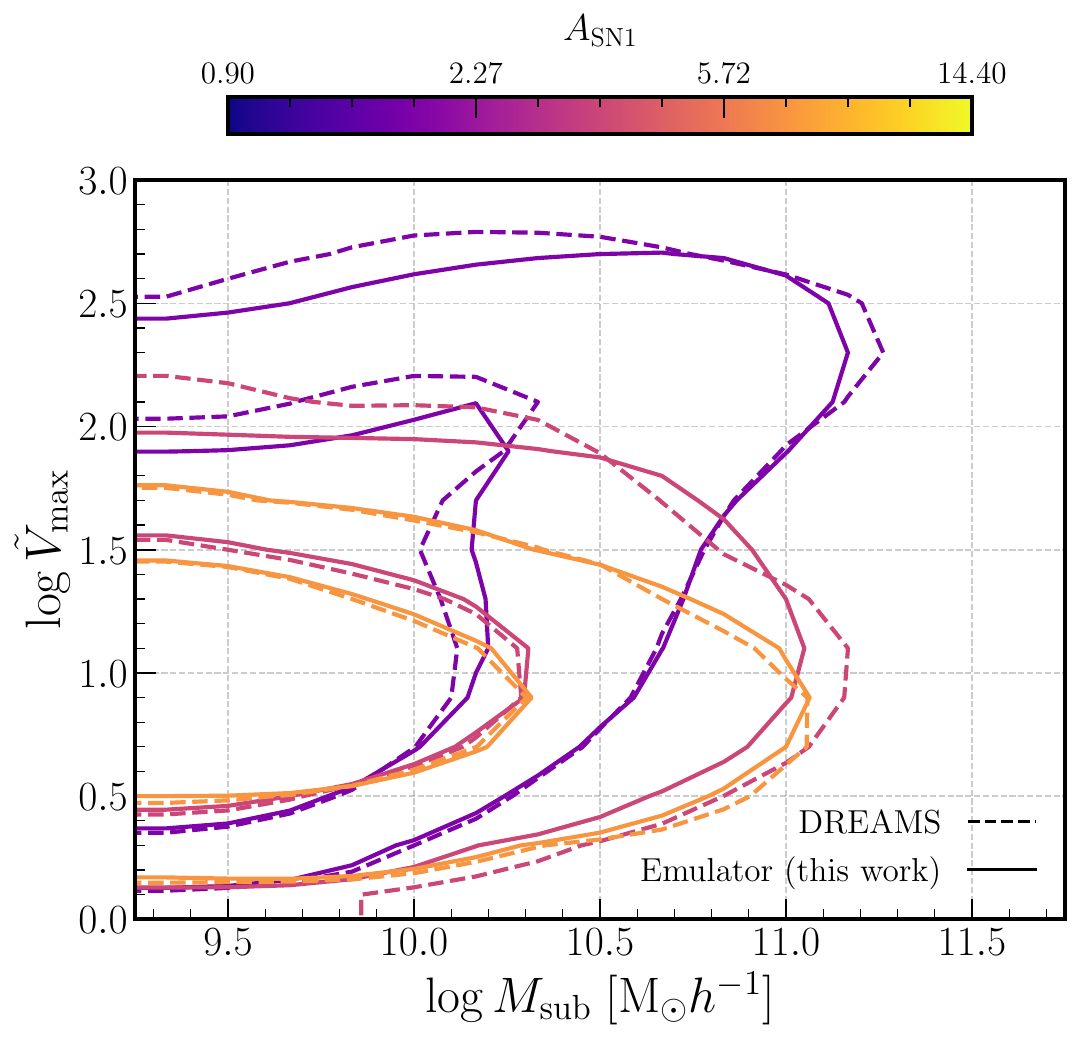}
    \caption{
    The 2-dimensional distributions of total mass and concentration proxy of satellites generated by \nn (solid) and extracted from the DREAMS simulations (dashed).
    The 68\% and 95\% contours are shown for the three bins of \sno.
    }
    \label{fig:mcvir_contour}
\end{figure}

The DM concentration describes the internal structure of a halo and is known to correlate strongly with both the formation time and the environment of the halo~\citep[e.g.,][]{1999MNRAS.303..685N, 2002MNRAS.331...98V, 2006ApJ...652...71W, 2015MNRAS.450.1521C}.
A key relation in galaxy formation studies is the concentration-mass relation, which links the DM concentration to the halo mass.
This will also provide an important validation test for predicting inter-galaxy properties with \nn. 

We employ a similar procedure with the SHMR to calculate the concentration-mass relations.
To summarize, for each halo, the median, 16th, and 84th percentiles of the \textit{concentration proxy}, \vmaxtilde, are calculated for \msubhalo bins of 0.5 dex width.
Then, for each bin of the simulation parameters, we estimate the average and standard error of these percentiles across all halos.
Note that as with central galaxies, the distribution of \vmaxtilde is bimodal and strongly correlates with the satellite stellar mass and SN feedback parameters, especially \sno.
However, the bimodality is much less prominent for the satellite subhalos, making the average percentiles acceptable summaries for the concentration-mass relations.

Figure~\ref{fig:mcvir} presents the concentration-mass relations.
As with the SHMR, we show the average medians, 16th and 84th percentiles in the top rows, the residuals of the medians in the middle rows, and the residuals of the scatters in the bottom rows.
Overall, the relations vary most significantly with \sno.
Additionally, the concentration proxy decreases with increasing WDM mass, suggesting that WDM halos are less concentrated than CDM halos. 
This further supports the notion that WDM halos form later, thus affecting the stellar mass as seen in Figure~\ref{fig:shmr}. 
We leave a more detailed investigation to future work.

The concentration-mass relations between \nn and the simulations agree remarkably well, with both the average medians and scatters agreeing within 1-2 error bars.
As with the SHMRs, the errors in estimating both the average medians and scatters of the simulations increase significantly at high \msubhalo due to the limited number of satellites at these masses.  

To demonstrate that \nn can capture the bimodality in the DM concentration proxy \vmaxtilde of satellites, we compare the 2-dimensional distributions of \vmaxtilde and \msubhalo.
Figure~\ref{fig:mcvir_contour} shows the 68\% and 95\% contours of the \vmaxtilde and \msubhalo distributions as a function of \sno.
We only show \sno since it has the most significant impact on the concentration, as can be seen in Figure~\ref{fig:mcvir}.
The distributions of \vmaxtilde and \msubhalo exhibit a clear bimodal trend, although this bimodality is less pronounced than what we observe in the central galaxies (Figure~\ref{fig:flows_nd}).
Overall, the contours predicted by \nn align well with those extracted from the simulations. 
However, the contours in the lowest \sno bin show less agreement, likely because the bimodality is more pronounced in this bin and thus requires more training samples to capture fully.
 
\subsubsection{Spatial distribution and clustering}
\label{section:res_position}

\begin{figure*}
    \centering
    \includegraphics[width=0.375\linewidth]{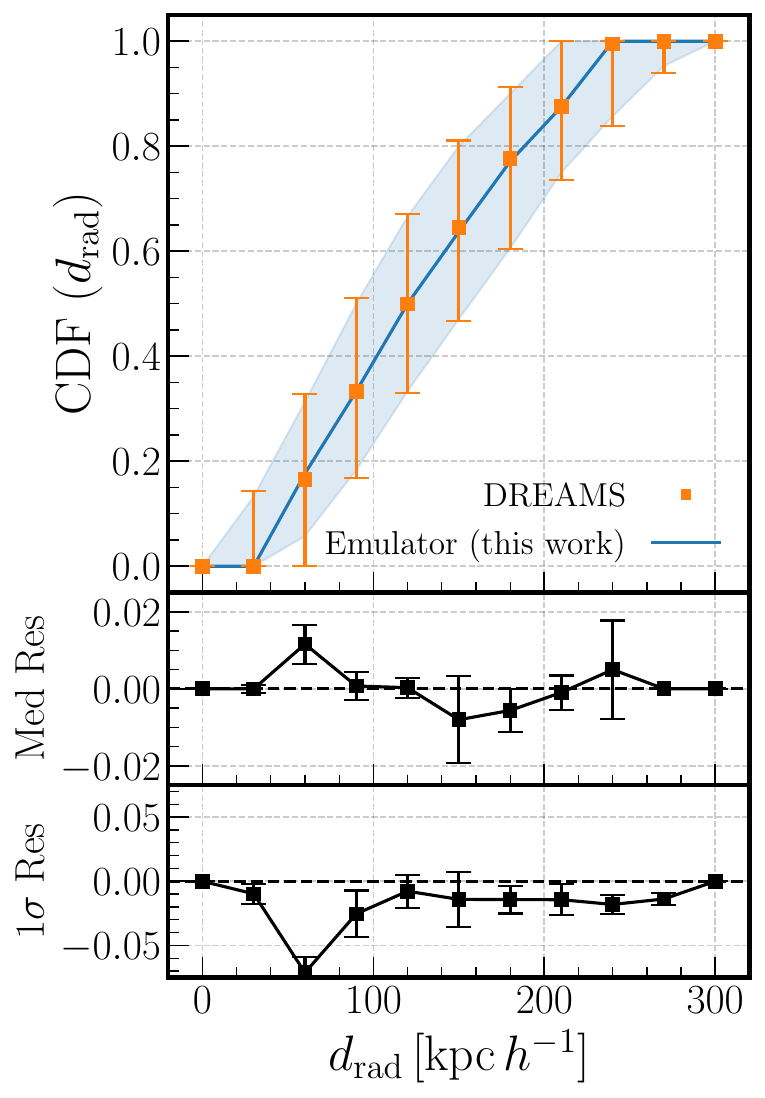}
    \includegraphics[width=0.375\linewidth]{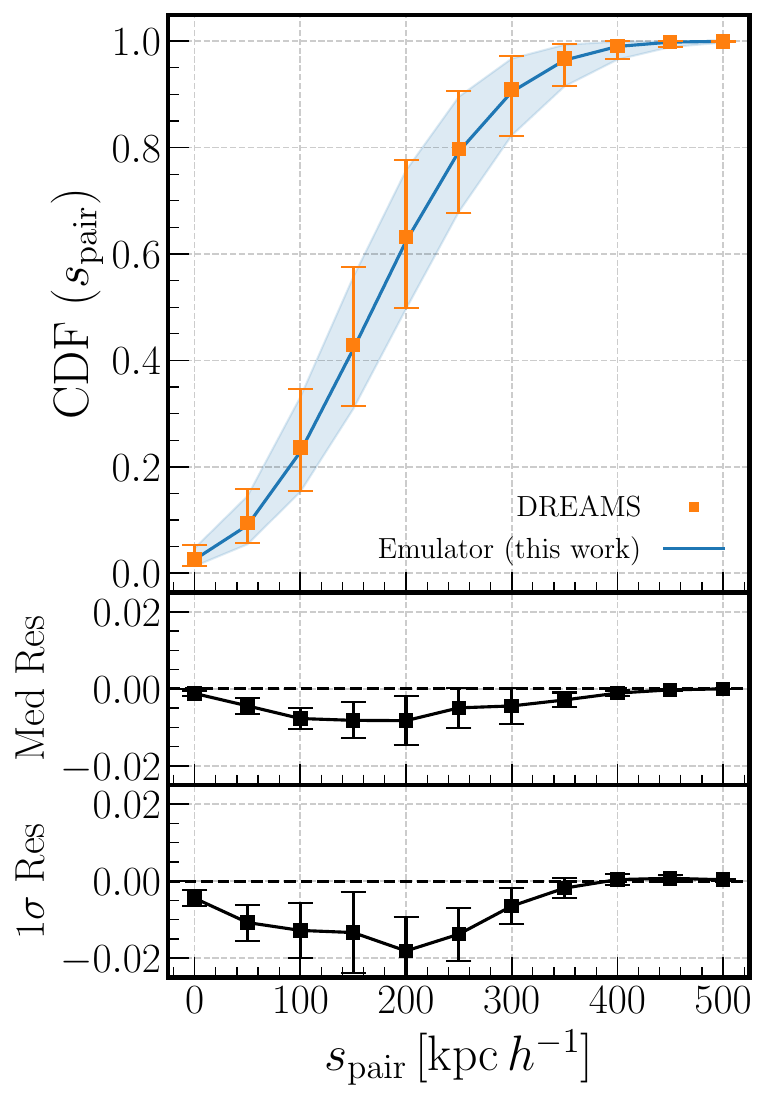}
    \caption{
    The cumulative distribution functions (CDFs) of radial distances (left) and pairwise separations (right).
    The top panels show the medians, 16th, and 84th percentiles of the CDFs for the \nn samples (blue band) and the simulations (error bars).
    The middle and bottom panels show the residuals between the \nn samples and the simulations of the medians and the $1\sigma$ intervals, respectively.
    }
    \label{fig:r_pairwise_cdf}
\end{figure*}

\begin{figure*}
    \centering
    \includegraphics[width=\linewidth]{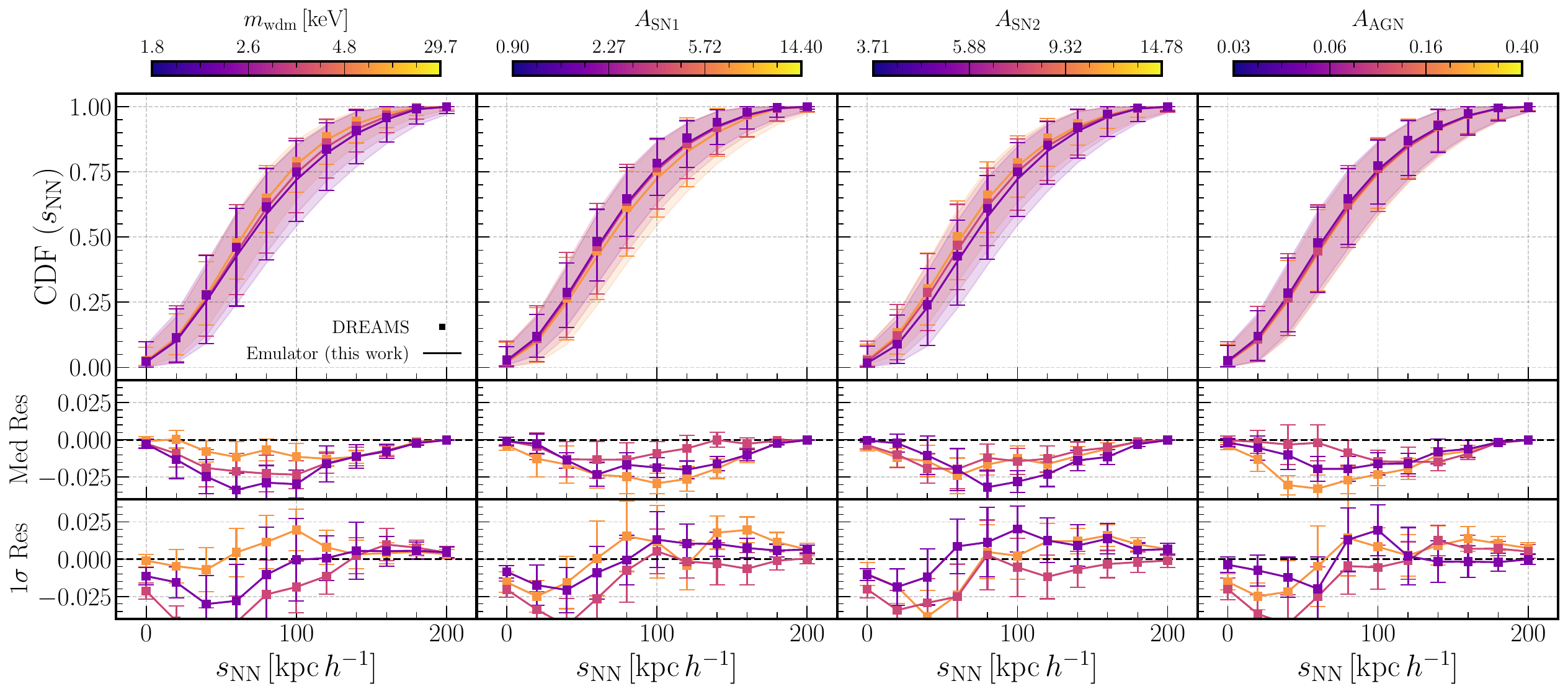}
    \caption{
    The cumulative distributions of the nearest-neighbor separations of satellites generated by \nn and extracted from the DREAMS simulations.
    Panels are the same as Figure~\ref{fig:shmr}. 
    }
    \label{fig:nearest_dist}
\end{figure*}

As discussed in \S\ref{section:method}, the \nn framework is unique in its point-cloud approach for modeling the phase-space of satellites. 
Point-cloud can resolve arbitrarily small spatial scales, down to the spatial resolution of the simulations, and thus are better than binning/voxelization for representing sparse data, including satellite distributions.
Here, we provide the validation tests for the spatial distribution and clustering of the satellites.

We first compute the radial distribution and the pairwise separation distribution of satellites generated by \nn and compare them to those of the simulations. 
The radial distribution is important for probing the underlying subhalo distribution~\cite[e.g.,][]{2005ApJ...618..557N}, the disruption of subhalo and satellites due to the central galaxy~\cite[e.g.,][]{2020MNRAS.491.1471S}, and the DM model~\cite[e.g.,][]{2021ApJ...920L..11N, 2021MNRAS.507.4826L}.
It is also a common statistic reported by satellite galaxy surveys~\citep[e.g.,][]{2020ApJ...902..124C, 2022ApJ...927..121W, 2024arXiv240414498M}, which provides a basis for comparing \nn predictions with the results of these surveys.
However, before any direct comparison can be made, we note that only \textit{projected} radial distributions are reported in these surveys. 
Additionally, sample selection effects need to be properly accounted for.
On the other hand, the pairwise separation distribution is a measure of the 1-halo clustering and is important for applications from creating mock galaxies catalogs for strong lensing~\citep[e.g.,][]{2023ApJ...942...75W, 2024arXiv240303253G} and galaxy surveys~\citep[e.g.,][]{2024MNRAS.532..903S, 2022A&A...658A..20H} to semi-analytic modeling of satellite distributions~\citep[e.g.,][]{2021MNRAS.502..621J, 2022MNRAS.509.2624G, 2023arXiv231105676F}.

Figure~\ref{fig:r_pairwise_cdf} shows the normalized, cumulative distribution of the radial distances $d_\mathrm{rad}$(left) and the pairwise separations $s_\mathrm{pair}$ (right).
We compute and display the median, 16th, and 84th percentiles of each distribution across \textit{all} halos in the \nn samples and the simulations. 
As with the mass functions, we estimate the percentiles and their errors using bootstrapping.
The top panels show the percentiles of the distributions, while the middle and bottom panels show the residuals between \nn and the simulations of the medians and $1\sigma$ intervals respectively. 

We do not show results for individual bins of simulation parameters \simparam, since we do not observe any significant variations in both the normalized radial distribution and pairwise separation distribution with respect to these parameters. 
This is despite differences in the stellar mass $M_\mathrm{cent, \star}$ and concentration proxy \vmaxtilde of the central galaxies between bins of \sno and \snt (see Figure~\ref{fig:flows}).
Note that the radial distribution is slightly more concentrated at higher values of \snt, though this effect is minor compared to the considerable scatter in the distribution, likely due to halo-to-halo variations.
Additionally, while the normalized distributions are relatively invariant to the simulation parameters, the \textit{un-normalized} distributions are not because the simulation parameters still influence the total number of satellites.

In general, we see a good agreement between the median distributions of the \nn samples and the simulations, with the residuals remaining with 1-2 error bars.
We find that \nn underestimates the $1\sigma$ intervals of the radial distributions.
Except for 1 radial bin at around $d_\mathrm{rad} \approx 50 \, \mathrm{kpc}$, the biases typically remain with 1-2 error bars and are small compared to the values of the $1\sigma$ intervals.

On the other hand, we observe a clear bias in the pairwise separation distributions at lower values of $s_\mathrm{pair}$. 
Specifically, \nn underestimates both the median and the $1\sigma$ interval of the cumulative distribution up to 1\% and 2\% respectively, with the largest discrepancy occurring at around $s_\mathrm{pair} \approx 200 \, \kpch$.
This indicates that while satellites generated by \nn have similar radial extensions to those in the simulations, they are typically less clustered. 
We quantify this discrepancy by computing the median and maximum $s_\mathrm{pair}$ between satellites in the same halo for the \nn samples and the simulations. 
On average, the median $s_\mathrm{pair}$ are $167.94 \pm 34.34 \, \kpch$ and $165.97 \pm 36.15 \, \kpch$ for the \nn and the simulations, respectively.
This corresponds to an overestimation by approximately $1\%$, although this effect is much smaller compared to the spread.
Likewise, the maximum $s_\mathrm{pair}$ are $372.35 \pm 78.02 \, \kpch$ and $361.32 \pm 83.98 \, \kpch$ for the \nn and the simulations, respectively, indicating an overestimation by approximately $3\%$.

To further investigate the spatial clustering, we identify the nearest neighbor (including the central galaxy) for each satellite in our samples and show the cumulative distribution of nearest-neighbor separations $s_\mathrm{NN}$ in Figure~\ref{fig:nearest_dist}.
Similar to Figure~\ref{fig:r_pairwise_cdf}, we show the medians, 16th and 84th percentiles in the top row, the median residuals in the middle row, and the $1\sigma$ interval residuals in the bottom row. 
The percentiles and their errors are estimated using the same bootstrapping procedure.
We briefly note that, in contrast to the radial distances and pairwise separations, the distributions of $s_\mathrm{NN}$ vary strongly with respect to \mwdm, \sno, and \snt. 
Thus, in Figure~\ref{fig:nearest_dist}, we show results for multiple bins of simulation parameters.
Additionally, this implies that the simulation parameters may influence the small-scale, local clustering of satellites more strongly than they do for the overall radial structure.

While \nn can capture the general trends in the distributions of $s_\mathrm{NN}$ with respect to the simulation parameters, we observe similar biases at small distances.
Across all parameter bins, \nn underestimates both the medians and $1\sigma$ intervals up to approximately 2-3\%, as evident from the middle and bottom rows. 
The largest biases occur at around $s_\mathrm{NN} \approx 50 \, \kpch$.
The biases in the $s_\mathrm{NN}$ distribution likely propagate into larger scales, resulting in biases in the pairwise separation $s_\mathrm{pair}$ distributions, as seen in Figure~\ref{fig:r_pairwise_cdf}.

As with the pairwise distances, we compute the median and maximum $s_\mathrm{NN}$ of satellites within the same halo in the \nn samples and the simulations.
The median values of $s_\mathrm{NN}$ are $93.40 \pm 29.6 \, \kpch$ and $90.72 \pm 31.43 \, \kpch$ for \nn and the simulations, respectively.
This corresponds to an overestimation by about 3\%, similar to the values obtained from the pairwise distances.
However, we find that the \nn overestimates the maximum $s_\mathrm{NN}$ by about 13\%, with \nn and simulation values of $230.95 \pm 61.91 \, \kpch$ and $203.48 \pm 49.56 \, \kpch$, respectively.
This further supports that while \nn can capture the overall radial extensions of satellites (as seen in the left panels of Figure~\ref{fig:r_pairwise_cdf}), it is unable to capture the local clustering between satellites. 
Lastly, we note that we also observe a similar bias in the velocity space, in which the relative velocities of \nn satellites are larger than seen in the simulations. 
We show this result in Appendix~\ref{app:res_velocity}.

To summarize, we examine the radial distribution, pairwise separation distribution, and nearest-neighbor distance distribution of satellites in the \nn halos and the simulations.
Overall, the shapes of distributions predicted by \nn generally match those in the simulations, including variations with respect to the simulation parameters \simparam. 
However, we find that while satellites generated by \nn match the radial extensions of those in the simulations, they are typically less clustered.
The median and maximum pairwise separations between satellites in the \nn halos are larger than the simulation values by about 1-3\%.
However, while the median nearest-neighbor separations between satellites in the \nn halos are larger by about 3\%, the maximum nearest-neighbor separations can differ by 13\%.
We thus conclude that \nn cannot yet capture the local clustering between satellites and caution readers against applying \nn in scenarios where accurate modeling of local clustering is important.
We further discuss the current limitations of our model and potential ways to address them in future work in \S\ref{section:discussion}.

\section{Discussion}
\label{section:discussion}

\subsection{Comparison with traditional methods}
\label{section:discussion_traditional}

Generative models such as \nn can be considered non-parametric techniques for modeling the galaxy-halo connection. 
Compared to the traditional techniques such as HOD, subhalo abundance matching (SHAM), and semi-analytic models (SAM), \nn does not make explicit assumptions about the distributions of halo and galaxy properties. 
Instead, it learns these distributions directly from simulations, where the underlying assumptions are embedded implicitly.
In contrast, HOD assumes a specific form for the galaxy-halo relationship, typically relying on a parametrized model for the number of galaxies as a function of halo mass~\citep{1952ApJ...116..144N}. 
SHAM assumes a monotonic relationship between a galaxy property and a corresponding halo property~\citep[e.g.,][]{2004ApJ...609...35K, 2004MNRAS.353..189V, 2006MNRAS.371.1173V, 2006ApJ...647..201C, 2010ApJ...717..379B, 2010ApJ...710..903M, 2011ApJ...742...16T, 2013ApJ...771...30R, 2016MNRAS.460.3100C}.
SAMs rely on detailed analytic prescriptions for various physical processes, such as gas cooling, star formation, and feedback, making assumptions about how these processes operate and interact over cosmic time within the framework of hierarchical structure formation~\citep[e.g.,][]{1994MNRAS.271..781C, 2000MNRAS.319..168C, somerville2015, 2019MNRAS.483.2983Y, 2021MNRAS.503.3698H, 2021MNRAS.506.4011E}.
Below we discuss the key advantages of \nn:
\begin{itemize}
    \item The use of neural networks for non-parametric modeling allows \nn to learn more complex distributions. 
    This is especially important for modeling non-Gaussian features in target distributions that are otherwise challenging to capture.
    For example, both the flows and the VDM can capture the bimodality in the DM concentration proxy \vmaxtilde of the central and satellite galaxies, as shown in Figure~\ref{fig:flows_nd} and Figure~\ref{fig:mcvir_contour} respectively.
    
    Additionally, we show that \nn can capture second-order statistics, such as the scatters in the target distributions and relations. 
    Second-order statistics are important because they (1) are observationally quantifiable~\citep{2019NatCo..10.2504F}, (2) contain information in~\citep[e.g.,][]{2018MNRAS.474.4089T, 2018MNRAS.478.2618F, 2020MNRAS.493.1361F, 2022ApJ...933...48F, 2020MNRAS.495..686A, Hadzhiyska2021, 10.1093/mnras/stac2297}, and (3) are a source of systematics in cosmological inference~\citep{2024MNRAS.530.3127Z}.
    Although we have primarily validated the scatters in the distributions and relations in this work, future studies will extend this analysis to include additional important statistics, such as correlations between the scatters of different properties~\citep[e.g.,][]{2020MNRAS.493.1361F, 2020MNRAS.495..686A}.
    As \nn does not make explicit assumptions about the underlying distributions, extending this analysis is both natural and straightforward.
    
    \item
    The VDM and the normalizing flows employed in \nn can model high-dimensional distributions, as they learn directly from the data without requiring explicit assumptions for each parameter. 
    Parametric methods often become less efficient and more complex as the number of parameters increases.
    Additionally, each new parameter typically necessitates additional modeling assumptions. 
    The traditional HOD method does not incorporate information other than the halo mass.
    SHAMs assume a monotonic relationship between a galaxy property and a halo property, which can oversimplify the underlying physics, particularly in high-dimensional spaces where additional factors like subhalo disruption and galaxy mergers play a significant role.
    SAMs, while more flexible, require re-calibration whenever new parameters or physical processes are introduced.
    Additionally, modeling subhalos in both SAMs and SHAMs require full merger trees, which can be expensive to generate.
    In contrast, \nn scales naturally with dimensionality, efficiently handling larger parameter spaces by modeling complex, data-driven relationships.

    \item \nn can be readily extended to different DM scenarios and baryonic physics, provided that corresponding simulations are available. 
    Unlike traditional parametric techniques, which require specific assumptions and models tailored to each scenario, \nn learns directly from the simulations, allowing it to adapt to a wide range of physical conditions without the need for reparametrization. 
    For example, HODs and SHAMs would require adjustments to their underlying models. 
    SAMs need to be re-calibrated, which involves re-tuning a large number of parameters that govern various physical processes.
    This can be time-consuming, as it requires extensive comparisons with observational data to ensure that the model predictions remain consistent across different scales and epochs.
    In comparison, re-training \nn simply requires feeding it new data corresponding to the updated simulation scenarios and potentially a new set of output/input features. 
    This flexibility makes \nn particularly advantageous for exploring complex or new models that are challenging to capture with traditional methods.

    \item SHAMs and SAMs assign galaxies onto DM halos in N-body simulations, which can introduce systematic biases when comparing results with those from hydrodynamic simulations. 
    For example, halo finders may lose track of DM subhalos in N-body simulations due to tidal stripping or numerical resolutions, even though the galaxies residing within these DM subhalos may still survive.
    These galaxies, known as "orphan galaxies," can introduce significant systematics into SHAMs and SAMs, which often struggle to accurately track them~\citep{2008MNRAS.391.1489K, 2014MNRAS.437.3228G, 2017MNRAS.469..749P}.
    Although the effects of orphan satellites can be quantified and somewhat accounted for in SHAMs and SAMs~\citep[e.g.,][]{2007MNRAS.375....2D, 2012NewA...17..175B, 2014MNRAS.439..264G, 2022MNRAS.510.2900D, 2024MNRAS.532L..61H}, this requires additional modeling and assumptions.
    Therefore, it is advantageous to learn directly from hydrodynamic simulations.
    \nn follows this approach by generating galaxies directly from hydrodynamic simulations, with their properties conditioned on halo characteristics that are less susceptible to systematics related to tidal stripping and numerical resolution.
    
\end{itemize}

Lastly, we note that machine learning emulators are differentiable, i.e. their outputs change smoothly and predictably with respect to their inputs.
In the context of simulations, this allows for more efficient optimization and integration with gradient-based techniques and can significantly improve downstream tasks such as parameter inference. 
Building an end-to-end differentiable simulation, whether neural network-based or otherwise, is a growing area of research in astrophysics and cosmology~\citep[e.g.,][]{2021OJAp....4E...7H, 2021A&C....3700505M, 2021arXiv210412864M, 2024ApJS..270...36L, 2024AAS...24335401K}.
Both the flows and the VDM in \nn can individually be considered differentiable.
However, \nn as a whole is not end-to-end differentiable, since the computation of the number of satellites \nsat involves a rounding operation (as the flows predict $\log \nsat$ in practice, see \S\ref{section:dataset}).
In future work, we will experiment with alternative approaches to achieve full end-to-end differentiability.

\subsection{Comparison with other generative models}
\label{section:discussion_ml}

Similar non-parametric, machine learning-based techniques have also been employed in various aspects of galaxy formation and cosmology~\citep[e.g.,][]{2023mla..confE..21L, 2023arXiv230805145N, 2024arXiv240311692B, 2024arXiv240310609L}.
Here, we highlight the key advantages of \nn:
\begin{itemize}
    \item Existing emulators have employed techniques such as the Gaussian Process for emulating summary statistics such as satellite counts, and mass functions. 
    In contrast, \nn uses a Transformed-based VDM that can model individual satellites, from which we can extract summary statistics as shown in \S\ref{section:result}.
    This approach provides access to more detailed information, enabling a wider range of applications, such as generating mock subhalo catalogs, which summary statistics alone cannot offer.

    \item As discussed in \S\ref{section:method}, we represent satellite subhalos as point clouds.
    In contrast to binning and voxelization, point clouds carry more information about the data and can resolve smaller spatial scales, down to the resolution of the simulations.
    While convolutional neural networks (CNNs) have been successfully used for HOD applications, as shown in~\citealt{2024arXiv240800839B}, it is unclear if they can effectively model the small spatial scales and sparse spatial distributions addressed in this work.
    Accurately modeling small scales with CNNs would require decreasing the grid size, which significantly increases computational cost and can be challenging to handle with sparse data.

    \item As also discussed in \S\ref{section:vdm}, using VDMs for learning point-cloud data has notable advantages compared to other generative machine learning models.
    Compared to GANs, VDMs are more stable and less prone to mode collapse.
    They also provide a tractable way to evaluate the likelihood, making them suitable for inference and anomaly detection tasks.
    Additionally, VDM is more expressive than VAE and normalizing flows~\citep[e.g.,][]{2021arXiv210602808H, 2022arXiv220811970L}.
    Compared to normalizing flows, VDMs also scale better with high-dimensional data. 
\end{itemize}

\subsection{Current limitations and future outlook}
\label{section:discussion_limitations}

We now discuss the current limitations and future prospects of the \nn framework.
One significant limitation of \nn is its inability to fully capture the local clustering of satellites. 
While the radial distributions of satellites generally agree, satellites generated by \nn are typically less clustered than those in the simulations, as shown in \S\ref{section:res_position}.
Below we provide a few potential explanations and solutions.

\begin{itemize}
    \item A potential explanation is a phenomenon in machine learning literature known as the ``spectral bias'', where coordinate-based neural networks have difficulty learning high-frequency functions.
    During training, these networks tend to fit the low-frequency components of a target before the high-frequency components~\citep{2018arXiv180608734R, 2019arXiv190600425B, 2020arXiv200304560B}.
    Recent studies demonstrate that using Fourier features can assist networks in learning high-frequency components~\citep{2018arXiv180608734R, 2020arXiv200610739T}, including 1-dimensional time or position coordinates~\citep{2019arXiv191112864X, 2019arXiv190705321M, vaswani2023attention}, and more recently, 3-dimensional coordinates~\citep{2019arXiv190905215Z, 2020arXiv200308934M}.
    We will explore this approach by employing encoding layers that map the input positions and velocities of satellites onto the Fourier space. 
    Note that this is similar to the sinusoidal encoding layer used for the 1-dimensional diffusion time step in our Transformer (described in \S\ref{section:vdm}). 
    However, extending this to 3-dimensional positions and velocities is non-trivial, so we leave this for future work. 
    
    \item 
    In our setup, we have conditioned the halo and central galaxy flows only on the WDM mass and astrophysical parameters, without incorporating other factors. 
    We expect that environmental factors will also play a significant role. 
    Similarly, the VDM is conditioned on the properties of the halos and central galaxies but does not currently include aspects such as the morphology of the central galaxies or environmental information. 
    Although the DM concentration proxy \vmaxtilde of the central galaxy is expected to correlate with the environment and formation histories, explicitly including these factors in future work could improve the model's ability to capture local clustering more accurately.

    \item 
    Additionally, satellites with a nearest neighbor closer than $50 \, \kpch$ might have been recently or are currently being tidally disrupted.
    This disruption can affect \textsc{subfind}'s ability to accurately identify and calculate their properties~\citep[e.g.,][]{2011MNRAS.415.2293K, 2012MNRAS.423.1200O, 2024ApJ...970..178M}, resulting in noisier features that impact the training process.

    \item Lastly, even with data augmentation, a training dataset of 1,000 simulations might be too small for the VDM to fully model the 9-dimensional feature space for all satellites. 
    We expect more training samples from future simulations will reduce this bias, although the extent of this improvement is currently unclear.
    This is potentially another major caveat of the current model -- we do not know a priori how many simulations are necessary for the distributions of modeled properties to be satisfactory.
    Studying the scaling behavior of downstream properties (e.g., loss function, summary statistics) with the number of training simulations can help answer this in future work.
    
\end{itemize}

In addition, we outline the general limitations of \nn:

\begin{itemize}

\item Our model is conditioned on the properties of the halo and central galaxy from the hydrodynamic simulation. 
Ideally, we would like to condition our model on the properties of the halo and central subhalo of the corresponding N-body simulation. 
We note that in this case, the positions of the halos in the N-body and hydrodynamic simulations may be different, so our model should also predict the displacement vector. 
We leave this for future work.

\item As discussed in \S\ref{section:sim_ic}, the halos in the TNG-WDM suite are constrained to a relatively narrow mass range, similar to that of the Milky Way.
In addition, to manage the computational costs of simulating dense environments, these halos are selected based on an isolation criterion (i.e., they are not close to halos with virial mass greater than $7.2 \times 10^{11} \, \modoth$).
For other applications, we will train the models on halos with much wider variations in mass and environment to fully exploit our method's capabilities. 

\item As discussed previously, although both the flows and the VDM are differentiable, the end-to-end \nn framework is not, due to the rounding operator when calculating \nsat.
Future work will explore alternative approaches to achieve full end-to-end differentiability.

\item Although we apply data augmentation techniques to incorporate rotational equivariance, our model does not explicitly impose this symmetry.
Building such symmetry into our model may further improve its accuracy, performance, and simulation efficiency. 
Additionally, although not necessary for this work, in future work, we will explore incorporating E(3) symmetry, i.e. translation and rotational symmetry.
This is important when there is a positional offset between the halo and central galaxy, which is expected if we condition our model using halos from N-body simulations (instead of hydrodynamic simulations), as discussed in the previous point.

\item As discussed in \S\ref{section:method_overview}, the goal of this paper is to highlight the use of VDM and point clouds for modeling satellite galaxies.
To simplify the analysis, we combine the properties of halos and central galaxies into the same normalizing flows, effectively modeling the \textit{joint distribution} of halos and central galaxies. 
In future work, it is better to separate them and instead model the \textit{conditional distribution} of central galaxies given the halos. 
An interesting avenue for exploration is using a hierarchical normalizing flows approach, similar to what is described in~\cite{Lovell_2023}.
Specifically, we can employ one flow to model the conditional distribution of halos given the simulation parameters, and another to model the conditional distribution of central galaxies given the halo and simulation parameters.
This can be helpful for applications in which we already have access to the halos but not the central galaxies, e.g., painting galaxies onto existing halos in N-body simulations.
\end{itemize}

\section{Conclusion}
\label{section:conclusion}

In this work, we develop an emulator for Milky Way-mass halos, and their central and satellite galaxies by using normalizing flows and variational diffusion models (VDM). 
Our framework, \nn, is hierarchical: 
\begin{enumerate}
    \item The normalizing flows capture the conditional distribution of halo and central galaxies given the simulation parameters.
    \item The VDM then learns the properties of satellite galaxies given the halo, central galaxy, and simulation parameters. 
\end{enumerate}
This approach allows us to capture the complex dependencies between different levels of structure in the simulations.

We represent satellite galaxies in each halo as a point cloud centered around the halo center. 
This eliminates the need for binning/voxelization and allows simulation details at fine spatial scales (up to the resolution often simulations) to be captured. 
The VDM employs a Transformer-based score model, which leverages self-attention mechanisms, to effectively model this point cloud.

We train the \nn framework on the TNG-WDM simulation suite of the DREAMS project, which consists of 1024 simulations with varying WDM mass and astrophysical parameters, particularly the strength of baryonic feedback from supernova and active galactic nuclei.
We generate a test dataset by sampling halos and central halos using the flows, and then satellite galaxies using the VDM.
Given the WDM mass and astrophysical parameters, the flows sample key properties such as the virial mass and stellar mass of the halos, as well as the total mass, stellar mass, DM concentration proxy, and velocity offset for the central galaxies.
The VDM then takes the generated halos and central galaxies, along with the simulation parameters, to generate satellite galaxies, including the positions, velocities, halo and stellar mass, and DM concentration of each satellite.    
We summarize our main findings below:
\begin{itemize}
    \item
    In \S\ref{section:res_halo_prop} and Appendix~\ref{app:res_flows}, we compare the halos and central galaxies generated by \nn with those in the simulations. 
    Overall, the flows effectively capture the distributions of halo and galaxy properties, along with their complex dependencies on the simulation parameters. 
    The edges of the distributions tend to show less agreement, which is expected given the limited number of training samples. 
    Most notably, both the DM concentration proxy and the velocity offset display strong non-Gaussian features; the concentration proxy is bimodal and strongly correlates with the halo stellar mass, while the velocity offset has a long tail extending toward high velocities. 
    Despite the complexities of these distributions, the flows accurately capture them.
    
    \item 
    In \S\ref{section:res_sat_prop}, we compare key summary statistics between \nn satellites and the simulations.
    We demonstrate that \nn can capture the satellite stellar mass functions, stellar-to-halo mass relations, and concentration-mass relations accurately.
    Notably, \nn can capture both the medians and the scatters of the target distributions and relations. 
    Additionally, we show that \nn can model complex relations such as the bimodal distributions of the concentration proxy \vmaxtilde of satellite galaxies (Figure~\ref{fig:mcvir_contour}). 
    However, similar to the flows, the model tends to struggle towards the extreme values (e.g., high mass range), where there are fewer training samples to learn from. 
    Nonetheless, we note discrepancies between the \nn samples and simulations are generally within 1-2 error bars and are typically much smaller than the intrinsic spreads from halo-to-halo variations.

    \item
    In \S\ref{section:res_position}, we examine the spatial distribution of satellites, including the radial distribution, pairwise separations, and nearest-neighbor separations.
    The overall trends predicted by \nn are consistent with the simulations.
    However, we find that although \nn satellites have similar radial distributions to those from the simulations, they are typically less clustered than those in the simulations.
    We observe similar discrepancies in the velocity space (Appendix~\ref{app:res_velocity}), where \nn satellites tend to move faster relatively to each other compared to those in the simulations.
    We discuss the limitations of the current iteration of \nn and potential ways to overcome them in \S\ref{section:discussion}.
    For now, we caution readers against over-interpreting results when using \nn for applications in which accurate modeling of small-scale clustering and velocity distributions is crucial.

    \item Although this work is primarily on the methodology and validation of \nn, emulating halos and galaxies under alternative DM scenarios and baryonic feedback models has significant implications for constraining DM properties and understanding the broader impact of these factors on galaxy formation and evolution.
    \nn will enable studies of the intricate interaction between the effects of new physics and astrophysics at both \textit{the field as well as summary levels}.   
    During our analysis, we incidentally uncovered several interesting and potentially unexpected correlations between astrophysical and DM parameters while evaluating summary statistics.
    
    In \S\ref{section:res_halo_prop}, we observe that the DM concentration proxy \vmaxtilde of central galaxies follows a bimodal distribution, which may be linked to the bimodality in the formation histories of their halos.
    We see a similar, albeit less prominent, trend in \vmaxtilde of satellite galaxies in \S\ref{section:res_cvir}.
    In \S\ref{section:res_shmr},  when examining the stellar-to-halo mass relations, we find that below approximately $10^{10} \, \modoth$ total mass, WDM satellites generally have higher stellar masses than their CDM counterparts for the same DM mass, possibly due to WDM halos forming later, leading to delayed star formation.
    In \S\ref{section:res_cvir}, when examining the concentration-mass relations, we find that WDM satellites typically have lower concentrations at the same total mass, likely due to the later formation times of WDM halos.
    As we continue to improve \nn and further explore the complex interplay between DM and astrophysical parameters, addressing these correlations and understanding their implications---while also assessing whether they are numerical artifacts---will be a key priority of the DREAMS project. 
\end{itemize}

Overall, the \nn framework offers a novel approach for non-parametric modeling of the galaxy-halo connection at the field level, all while operating at a computational cost that is several orders of magnitude lower than that of hydrodynamic simulations.
We further emphasize that \nn can model individual satellites and complex relationships between halos, galaxies, and the simulation parameters, while not making any explicit assumptions about the underlying distributions.
Instead, \nn directly learns from the simulations, where the physical assumptions are embedded implicitly.
This allows \nn to capture more complex and high-dimensional distributions, as well as offering more flexibility to extend to different DM and baryonic physics scenarios, as discussed in \S\ref{section:discussion}.

We plan to utilize \nn for generating mock galaxy catalogs for applications such as galaxy clustering, weak and strong lensing, and potentially intensity mapping. 
For example, in a follow-up project, we will use \nn to generate strong lensing catalogs and compare our model against traditional methods such as SAM, SHAM, and HOD, particularly in the context of galaxy clustering at the field level.
We envision \nn to be especially valuable in applications where populating halos with galaxies is crucial. 
Additionally, since both the flows and VDM components of \nn allow efficient evaluation and sampling of the likelihood function, they hold potential for use in cosmological parameter inference and even outlier detection, though further tests are needed.

In conclusion, \nn offers a powerful and flexible framework for exploring various DM models and astrophysical scenarios at the field level.
Its ability to model complex, high-dimensional distributions and flexibility makes it a promising tool for a wide range of astrophysical and cosmological applications. 
As we continue to develop and improve \nn, we anticipate its application in even more areas, from generating mock catalogs to conducting cosmological parameter inference.

\software{
\texttt{corner}~\citep{corner},
\texttt{IPython}~\citep{PER-GRA:2007},
\texttt{jax}~\citep{jax2018github},
\texttt{Jupyter}~\citep{Kluyver2016JupyterN},
\texttt{Matplotlib}~\citep{Hunter:2007},
\texttt{NumPy}~\citep{harris2020array},
\texttt{SciPy}~\citep{2020NatMe..17..261V},
\texttt{Pandas}~\citep{reback2020pandas},
\texttt{PyTorch}~\citep{NEURIPS2019_9015}, 
\texttt{PyTorch Lightning}~\citep{william_falcon_2020_3828935},
\texttt{zuko}~\citep{rozet2022zuko}
}

\section*{Data Availability}

The code used for this article is available at \url{https://github.com/trivnguyen/nehod_torch}, which includes example tutorials and trained models (both the flows and VDM).
Only a subset of the simulation data, in the form of the \textsc{FoF/subfind} catalog, is included. 
For more information on the DREAMS simulation suites and full data access, please visit at \url{https://dreams-project.readthedocs.io/en/latest/}.

\section*{Acknowledgments}

We thank Mariangela Lisanti, Justin Read, and Xiaowei Ou for many meaningful discussions and suggestions, and Lucy Reading-Ikkanda for assisting with Figure~\ref{fig:flow_chart}. 

The Center for Computational Astrophysics at the Flatiron Institute is supported by the Simons Foundation.
The computations in this work were, in part, run at facilities supported by the Scientific Computing Core at the Flatiron Institute, a division of the Simons Foundation.
The data used in this work were, in part, hosted on equipment supported by the Scientific Computing Core at the Flatiron Institute, a division of the Simons Foundation.
TN, SM, CC, and LN are supported by the National Science Foundation under Cooperative Agreement PHY-2019786 (The NSF AI Institute for Artificial Intelligence and Fundamental Interactions, \url{http://iaifi.org/}).
SM is partly supported by the U.S. Department of Energy, Office of Science, Office of High Energy Physics of U.S. Department of Energy under grant Contract Number  DE-SC0012567. 
This work was performed in part at the Aspen Center for Physics, which is supported by National Science Foundation grant PHY-2210452.
PT acknowledges support from NSF-AST 2346977.
JBM acknowledges support from NSF AST-2307354 and AST-2408637.
LN is also supported by the Sloan Fellowship, the NSF CAREER award 2337864, and NSF award 2307788.  FYCR acknowledges the support of NSF award 2327192.
KEK is supported by the National Science Foundation Graduate Research Fellowship under Grant No. DGE-2039656.
DAA acknowledges support by NSF grant AST-2108944, NASA grant ATP23-0156, STScI grants JWST-GO-01712.009-A and JWST-AR-04357.001-A, Simons Foundation Award CCA-1018464, and Cottrell Scholar Award CS-CSA-2023-028 by the Research Corporation for Science Advancement.



\bibliographystyle{aasjournal}
\bibliography{vdm_dreams} 




\appendix

\section{Normalizing flows}
\label{app:flows}

Normalizing flows model the distribution $p(\yy)$ of the data \yy by transforming it from a base distribution $p(\uu)$ via an invertible transformation. 
Let $f$ be an invertible transformation that maps a random variable \uu to \yy, i.e. $\yy = f(\uu)$.
Using the change of variables formula, we can write the $p(\yy)$ as:
\begin{equation}
    p(\yy) = p(f^{-1}(\yy)) \, \left| \det\left(\frac{\partial f^{-1}}{\partial \yy}\right) \right|
\end{equation}
where the last term is the absolute value of the determinant of the Jacobian matrix of $f^{-1}$ at \yy and accounts for the change in volume under the transformation.
The base distribution $p(\uu)$ is often set to a simple, well-understood distribution such as a standard Gaussian.
The transformation $f$ is a neural network with trainable parameters but with a few restrictions.
First, we emphasize the invertibility of the transformation, which allows for both sampling (forward transformation from $p(\uu)$ to $p(\yy))$ and density estimation (inverse transformation from $p(\yy)$ to $p(\uu)$).
In addition, the Jacobian determinant must be tractable to ensure the model can be efficiently trained and applied to high-dimensional data.
These requirements restrict the type of transformations available, with common flows being the masked autoregressive flows~\citep{DBLP:conf/icml/GermainGML15, 10.5555/3294771.3294994} and neural spline flows~\citep{2019arXiv190604032D}.

In practice, the normalizing flows are constructed from a sequence of $T$ discrete transformations $f = f_1 \circ f_2 \circ ... \circ f_T$, where each transformation $f_i$ is also invertible and has tractable Jacobian determinant.
The data likelihood $p(\yy)$ can be written as:
\begin{equation}
    p_{T}(\yy) = p_{0}(\uut{0}) \prod_{i=1}^{T} \left| \det\left(\frac{\partial f_i^{-1}}{\partial \uut{i-1}}\right) \right|
\end{equation}
where $\yy = \uut{T}$.
The optimization objective is thus simply the negative log-likelihood $\log p_{T}(\yy)$, i.e.
\begin{equation}
    \mathcal{L}_\mathrm{flows}(\yy) = -\log p_{0}(f^{-1}(\yy)) - \sum_{i=1}^{T} \log \left| \det\left(\frac{\partial f_i^{-1}}{\partial \uut{i-1}}\right) \right|. 
\end{equation}
Each transformation $f_i$ is parameterized by a neural network with parameters $\phi_i$.
During training, the full set of parameters $\phi = \{\phi_1, \phi_2, ..., \phi_T\}$ is optimized simultaneously.
In the case of a conditional generative model, with conditioning features \vtheta, we simply include \vtheta into each transformation, i.e. $f_i(\yy) \rightarrow f_i(\yy, \vtheta)$.

\section{Diffusion Model}
\label{app:vdm}

In this Appendix, we present a summary of the inner workings of variational diffusion models (VDM), following the discussion in~\cite{2021arXiv210700630K}.
For more detailed mathematical formalism and derivation, we also referred readers to a review in ~\cite{2022arXiv220811970L}.

\subsection{Forward diffusion}
\label{app:forward_diff}
In the context of a Gaussian diffusion model, forward diffusion refers to the process of gradually corrupting an original data \yy by incrementally adding Gaussian noise to it over $T$ discrete timesteps. 
Let the noisy version of \yy at some time step $t \in [0, 1]$ be the \textit{latent variables} \zzt{t}, we may define the \textit{variance-preserving} mapping from \yy to \zzt{t} (see~\cite{2015arXiv150303585S, 2020arXiv200611239H}) as:
\begin{equation}
    \label{eq:forward_diff}
    q(\zzt{t} | \yy) = \mathcal{N}\left(\zzt{t} | \alpha_t \yy, \sigma^2_t \mathbf{I}\right),
\end{equation}
where $\sigma^2_t$ is a strictly positive scalar-valued function of $t$ that represents the variance of the noise added at each step, and $\alpha_t = \sqrt{1 - \sigma^2_t}$.
Note that the distribution $q(\zzt{t} | \yy)$ is conditioned on the data \yy.

One can define a \textit{signal-to-noise ratio} $\mathrm{SNR(t)} \equiv \alpha^2_t / \sigma^2_t$.
As $t$ progresses from 0 (least noisy) to $1$ (most noisy), we require that $\sigma^2_t$ increases, indicating that more noise is being added, which in turn reduces the signal-to-noise ratio $\mathrm{SNR(t)}$.
With a sufficiently small $\mathrm{SNR(1)}$, the variance-preserving diffusion transformation ensures that the maximally noise-corrupted data at $t=1$ follow a standard Gaussian distribution, i.e.
\begin{equation}
    q(\zzt{1} | \yy) \approx \mathcal{N}\left(\zzt{1}| 0, \mathbf{I}\right).
\end{equation}
The condition offers key benefits, including analytical tractability, consistency in variance, and ease of sampling, which collectively enhance model stability and efficiency. 

As mentioned in Section~\ref{section:vdm}, the noise schedule, which controls how $\sigma^2_t$ depends on $t$, is the most critical part for the performance of diffusion models~\citep{2023arXiv230110972C}.
In our framework, we employ a linear noise schedule with the following functional form:
\begin{equation}
    \label{eq:noise_schedule}
    \sigma^2_t = \texttt{Sigmoid}\left(\gamma_\eta(t)\right), \quad \mathrm{where \;} \gamma_\eta(t) = \gamma_\mathrm{max} - (\gamma_\mathrm{max} - \gamma_\mathrm{min}) t,
\end{equation}
where $\eta = \{\gamma_\mathrm{max}, \gamma_\mathrm{min}\}$ is the trainable parameter.
It is straightforward to see that $\mathrm{SNR(t)} = \exp (-\gamma(t))$, so $\gamma_\mathrm{max}$ and $\gamma_\mathrm{min}$ also controls the minimum and maximum signal-to-noise ratio of the forward diffusion process. 
During our training, we initialize $\{\gamma_\mathrm{min}, \gamma_\mathrm{max}\}$ to be $\{-8, 14\}$, though there is no restriction of what values they can take.
Other possibilities for the noise schedule include, for example, a linear or cosine noise schedule with fixed hyperparameters, or using a monotonic neural network (in which case $\eta$ will be the parameters of the network).

\subsection{Reverse diffusion}
\label{app:reverse_diff}
To generate new data, we would like to revert the forward diffusion process in Equation~\ref{eq:forward_diff} by gradually denoising the corrupted data at $t=1$ over a series of discrete timesteps $T$.
This involves training a generative model that can sample a sequence of latent variables $\zzt{t}$ with time moving backward from $t=1$ to $t=0$.
Define the time steps to be $s(i) = (i-1)/T$ and $t(i) = i/T$ where $i \in [0, T]$, our generative model for data \yy can be written as:
\begin{equation}
    p(\yy) = p(\zz_{1}) p(\yy | \zz_{0}) \prod_{i=1}^T p(\zzt{s(i)} | \zzt{t(i)}).
\end{equation}
Familiar readers may recognize that this formalism is similar to a Markovian hierarchical variational autoencoder (VAE)~\citep{2016arXiv160604934K, 2016arXiv160202282K}, a generalization of VAEs that incorporates multiple levels of latent hierarchies and a Markov chain generative process.
Indeed, VDMs can be thought of as Markovian hierarchical VAEs where the latent dimension is the same as the data dimension and the latent encoding process is pre-defined as a linear Gaussian model (see~\cite{2022arXiv220811970L}).
This interpretation of VDMs is especially useful for understanding its optimization objective.

As stated in Appendix~\ref{app:forward_diff}, with sufficiently large $T$ and small $\mathrm{SNR(1)}$, we expect $q(\zzt{1})$ to be a standard Gaussian distribution.
We thus also model the first term $p(\zzt{1})$ as a standard Gaussian, i.e
\begin{equation}
    p(\zzt{1}) \approx \mathcal{N}(\zzt{1} | 0, \mathbf{I}).
\end{equation}
Similarly, as in the VAE analog, the second term represents the reconstructed data likelihood, which we will simply model as:
\begin{equation}
    p(\yy | \zzt{0}) \approx \mathcal{N}(\yy | \zzt{1}, \sigma^2\mathbf{I}).
\end{equation}
Note that $\sigma$ is a hyperparameter of the data likelihood and \textit{not} related to the noise schedule.
It determines how accurately the data is to be reconstructed during the training process, as well as the relative scaling between each term of the optimization objective (see Appendix~\ref{app:vdm_loss}).
In our framework, we set $\sigma = 0.001$.
Finally, we choose the last term to be:
\begin{equation}
    p(\zzt{s} | \yy) = q(\zzt{s} | \zzt{t}, \yy),
\end{equation}
where we have rewritten the ground truth $q(\zzt{s} | \zzt{t})$ to be conditioned on the target data $q(\zzt{s} | \zzt{t}, \yy)$.

\subsection{Variational optimization objective}
\label{app:vdm_loss}
To train the VDM, we minimize the negative Evidence Lower Bound (ELBO,~\cite{2013arXiv1312.6114K}),
\begin{align}
    - \log p(\yy) \leq -\mathrm{ELBO}(\yy) 
    = &- \mathbb{E}_{q(\zzt{1} | \yy)}\left[D_\mathrm{KL}\left(q(\zzt{1} | \yy) \| p(\zzt{1})\right)\right] \\ 
    &+ \mathbb{E}_{q(\zzt{0} | \yy)}\left[\log p(\yy | \zzt{0})\right] \\
    &+ \mathcal{L}_\mathrm{diff}(\yy),
\end{align}
where $D_\mathrm{KL}$ is the Kullback–Leibler (KL) divergence.
The first two terms are referred to as the \textit{prior matching loss} and the \textit{reconstruction loss}, paralleling their counterparts in the VAE analog.
The prior matching term minimizes the discrepancy between the final latent distribution and the Gaussian prior, while 
the reconstruction term ensures the accuracy between the data \yy and its reconstruction from the latent variable \zzt{0}.
It is worth noting that in a common setup where the noise schedule is linear, these two terms are ignored since they do not depend on the parameters of the noise prediction model (see~\cite{2020arXiv200611239H} for example).
However, as our noise schedule (Equation~\ref{eq:noise_schedule}) consists of trainable parameters $\eta$, we do not ignore these terms. 

Unlike in the VAE analog, the VDM loss includes an additional term, known as the \textit{forward-reverse consistency loss}, which can be written as:
\begin{equation}
    \label{eq:loss_diff}
    \mathcal{L}_\mathrm{diff}(\yy) = \sum_{i=1}^T \mathbb{E}_{q(\zzt{i} |  \yy)} D_\mathrm{KL} \left[q(\zzt{s(i)} | \zzt{t(i)}, \yy) \| p(\zzt{s(i)} | \zzt{t(i)}) \right]
\end{equation}
in case of finite $T$.
This term is minimized when the forward diffusion process $q(\zzt{s} | \zzt{t}, \yy)$ matches the reverse generative process $p_\varphi(\zzt{s} | \zzt{t})$ across \textit{all time steps}.

In practice, we parameterize the denoising model as a noise prediction model $\hat{\vect{\epsilon}}_\varphi(\zzt{t}, t)$, which is a neural network with parameters $\varphi$.
Additionally, the sum over the timesteps in Equation~\ref{eq:loss_diff} can be written as an expectation over the timesteps $t(i)$, which is sampled uniformly (i.e., over $i \sim U(1, T)$).
We can thus simplify Equation~\ref{eq:loss_diff} to be:
\begin{equation}
    \label{eq:loss_diff_discrete}
    \mathcal{L}_\mathrm{diff}(\yy) = 
    \frac{T}{2} \mathbb{E}_{\epsilon \sim \mathcal{N}(0, \mathbf{I}), i \sim U(1, T)}
    \left[w_\eta(t) \| \vect{\epsilon} - \hat{\vect{\epsilon}}_\varphi(\zzt{t}, t)\|^2_2
    \right],
\end{equation}
where the pre-factor $w(t)$ can be written as:
\begin{equation}
    w_\eta(t) = \exp(\gamma_\eta(t) - \gamma_\eta(s)) - 1.
\end{equation}
A step-by-step derivation of the diffusion loss can be found in~\cite{2022arXiv220811970L}.
Here, we note that rewriting $\mathcal{L}_\mathrm{diff}(\yy)$ as an expectation over the timesteps $t(i)$ allows for the negative ELBO through a Monte Carlo estimator.
This greatly increases the tractability and stability of the optimization process, as we do not need to simulate the entire trajectory from \yy to \zzt{0}.
Equation~\ref{eq:loss_diff_discrete} optimizes the parameter $\varphi$ of the noise prediction model and $\eta = \{\gamma_\mathrm{max}, \gamma_\mathrm{min}\}$ of the noise schedule simultaneously. 

Lastly, in case of a conditional generative process, we simply include the conditioning features \vtheta into the noise prediction model, i.e. $\hat{\vect{\epsilon}}_\varphi(\zzt{t}, t, \vtheta)$.

\subsection{Continuous-time diffusion}

The number of diffusion steps $T$ is an important hyperparameter of the VDM, with a higher $T$ generally leading to better performance (see Appendix F of~\cite{2021arXiv210700630K}).
In our framework, we employ a continuous-time VDM, corresponding to setting $T\rightarrow\infty$ and also reducing the number of hyperparameters by one. 
In this limit, the summation over the timesteps in Equation~\ref{eq:loss_diff} is simply replaced by an integration over $t$ from $t=0$ to $t=1$.
The final diffusion loss becomes:
\begin{equation}
    \label{eq:loss_diff_cont}
    \mathcal{L}_\mathrm{diff}(\yy) = 
    \frac{1}{2} \mathbb{E}_{\epsilon \sim \mathcal{N}(0, \mathbf{I}), t \sim U(0, 1)}
    \left[\gamma'_\eta(t) \| \vect{\epsilon} - \hat{\vect{\epsilon}}_\varphi(\zzt{t}, t)\|^2_2
    \right],
\end{equation}
where the time derivative $\gamma'_\eta(t) = d\gamma_\eta(t) / dt$ can be evaluated via automatic differentiation.
It is worth noting that in this limit, the diffusion process is governed by a stochastic differential equation; see~\cite{2020arXiv201113456S} for a more detailed discussion.

\section{Additional Results}

\subsection{Halo and central galaxies}
\label{app:res_flows}

\begin{figure*}
    \centering
    \includegraphics[width=\linewidth]{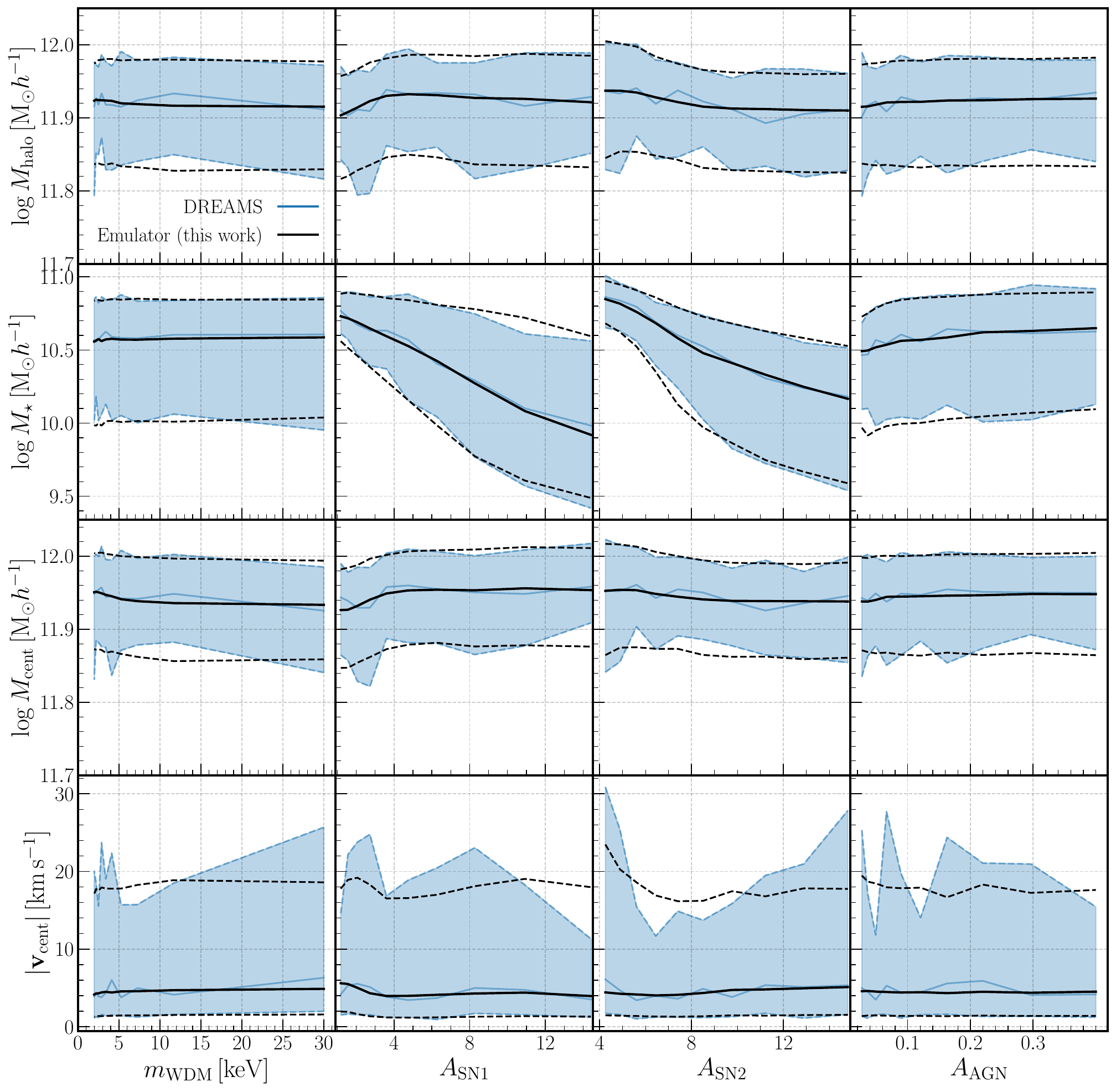}
    \caption{
    The properties of halos and central galaxies as a function of the simulation parameters \simparam (left to right).
    From top to bottom, the panels show the halo virial mass, halo stellar mass, central galaxy total mass, and the velocity offset between the halo and central galaxies.
    Panels are the same as in Figure~\ref{fig:flows}.
    }
    \label{fig:flows_all}
\end{figure*}

Figure~\ref{fig:flows_all} shows the rest of the output features of the normalizing flows, including distributions of the halo virial mass \mhalo, halo stellar mass \mstar, central galaxy total mass \mcent, and the velocity offset between the halo and central galaxies $|\mathbf{v}_\mathrm{cent}|$, as a function of the simulation parameters \simparam.

We do not observe strong variations of \mhalo and \mcent, as expected due to the mass selection criterion $(1.58-1.61) \times 10^{12} \, \mathrm{M_\odot}$, described in Section~\ref{section:sim_ic}.
However, since the target halo is selected from the initial DM-only, low-resolution uniform box, its mass could change when increasing the resolution and adding baryons in the final zoom-in simulation. 
Indeed, we observe minor but notable variations of \mhalo and \mcent to \sno and \snt.
These variations are captured by the flows and agree with the simulations.

The distribution of the halo stellar mass \mstar has similar trends to the that of the central galaxy, i.e. \mcentstar.
This is expected given the strong correlations between \mstar and \mcentstar (see Figure~\ref{fig:flows_nd}).
Similar to \mcentstar, we see that the distribution predicted by \nn generally align well with the simulations, with some discrepancies in the 16-84th percentiles. 
We expect the agreement will improve with more training samples.

Lastly, the distribution of the velocity offset $|\mathbf{v}_\mathrm{cent}|$ also shows good agreement between \nn and the simulations. 
The velocity offset $|\mathbf{v}_\mathrm{cent}|$ shows little variations with respect to the simulation parameters \simparam.
Notably, the distribution has a long tail extending toward higher values of $|\mathbf{v}_\mathrm{cent}|$.
We see from the slight disagreement in the 84th percentiles that the flows have some difficulties capturing this tail.
This is likely due to the limited number of training samples in this high-velocity range, and we anticipate improvement with additional training data.

\subsection{Satellite halo mass functions}
\label{app:mass_functions}

\begin{figure*}
    \centering
    \includegraphics[width=\linewidth]{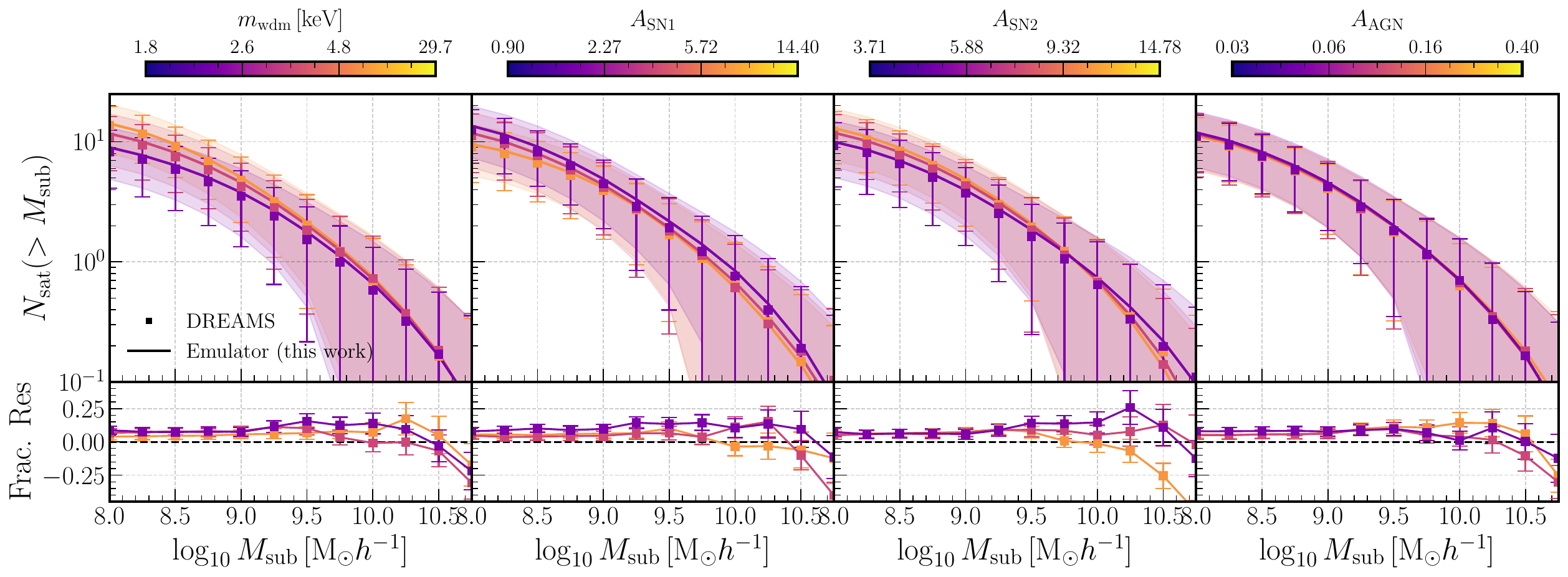}
    \caption{
    \textit{Top}: 
    The halo stellar mass functions (SHMFs) generated by \nn and extracted from the simulations.
    The columns show the variations of the SHMFs over the WDM mass \mwdm and astrophysical parameters \astroparam.
    In each column, the color denotes the SHMFs of a bin of the corresponding parameter. 
    The average SHMFs, along with the 1-$\sigma$ scatter, are shown as solid lines and shaded regions for \nn and error bars for the simulations.  
    \textit{Bottom}: 
    The fractional residuals of the SHMFs, estimated as the differences between \nn and the simulations divided by the simulation values.
    }
    \label{fig:mf_halo}
\end{figure*}

\begin{figure*}
    \centering
    \includegraphics[width=\linewidth]{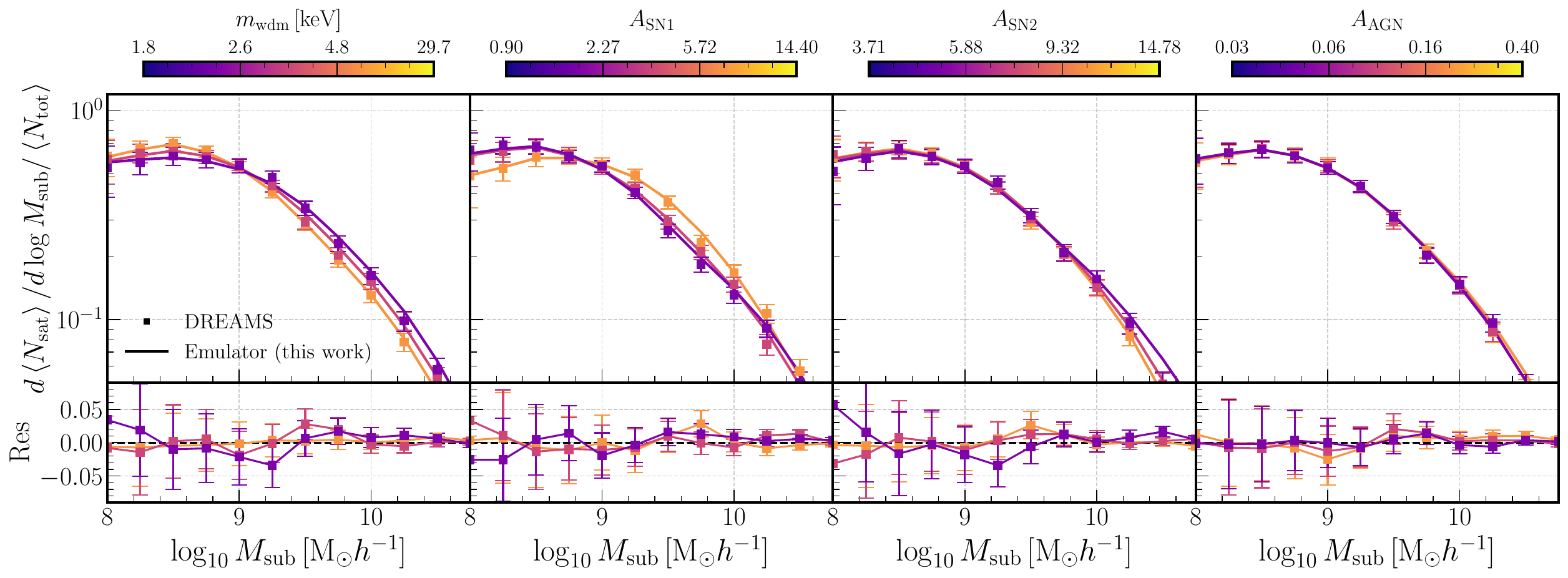}
    \caption{
    \textit{Top}: 
    The differential SHMFs $d \braket{\nsat} / d \log \msubhalo$, normalized by the average total number of satellites $\braket{N_\mathrm{tot}}$.
    The columns show the variations of the SHMFs over the WDM mass \mwdm and astrophysical parameters \astroparam.
    The \nn and simulation differential SHMFs are shown as solid lines and squares, respectively. 
    The error bars denote the standard errors of the simulations. 
    \textit{Bottom}: The residuals between \nn and simulation differential SHMFs.
    }
    \label{fig:mf_halo_grad}
\end{figure*}

Figure~\ref{fig:mf_halo} and Figure~\ref{fig:mf_halo_grad} show the satellite halo mass functions (SHMFs) and their differential forms, respectively. 

\subsection{Velocity distribution of satellites}
\label{app:res_velocity}

\begin{figure*}
    \centering
    \includegraphics[width=0.325\linewidth]{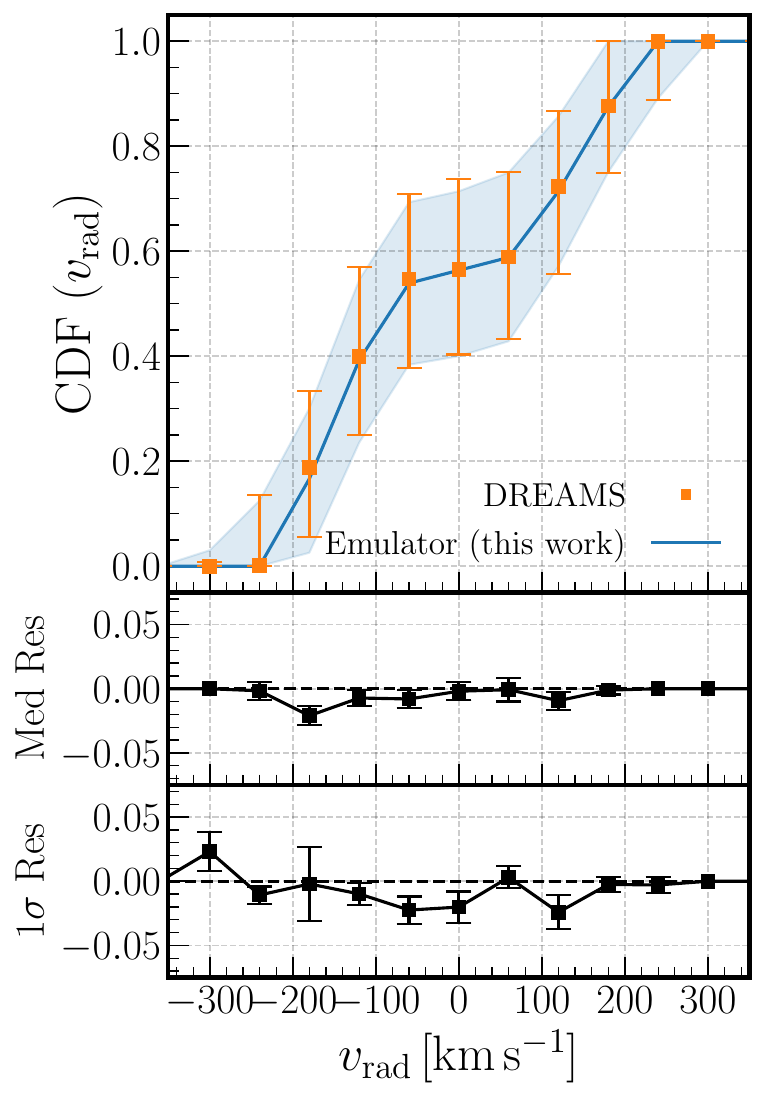}
    \includegraphics[width=0.325\linewidth]{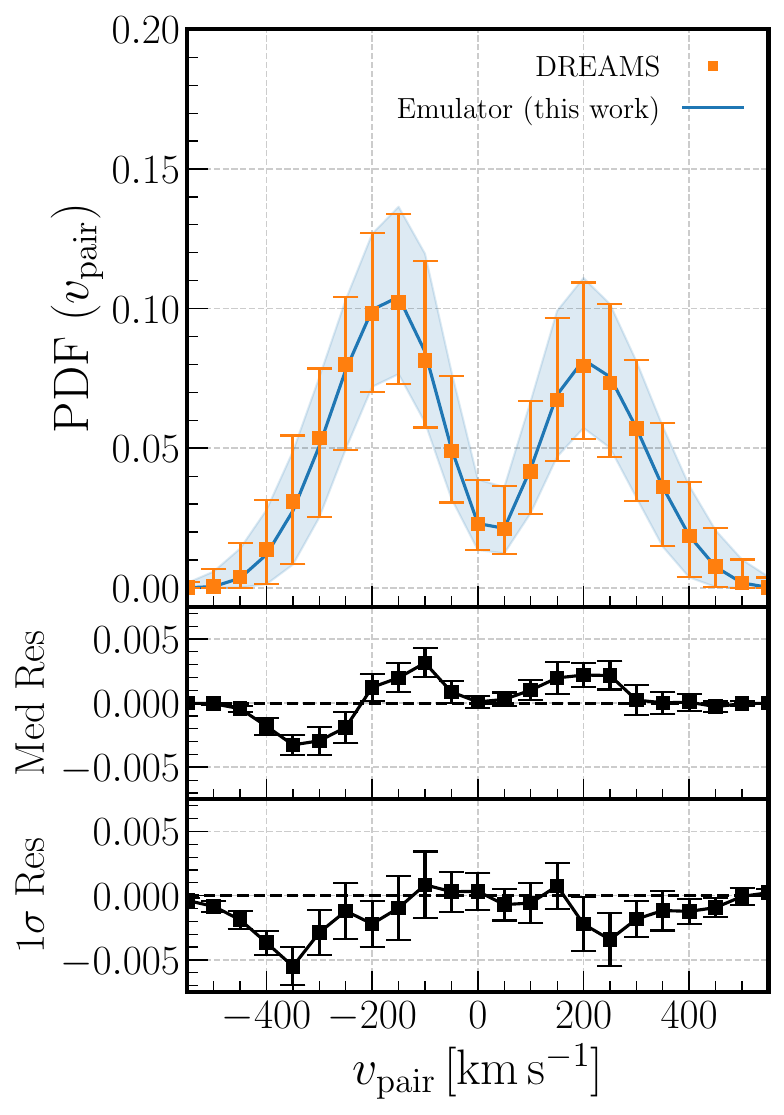}
    \includegraphics[width=0.325\linewidth]{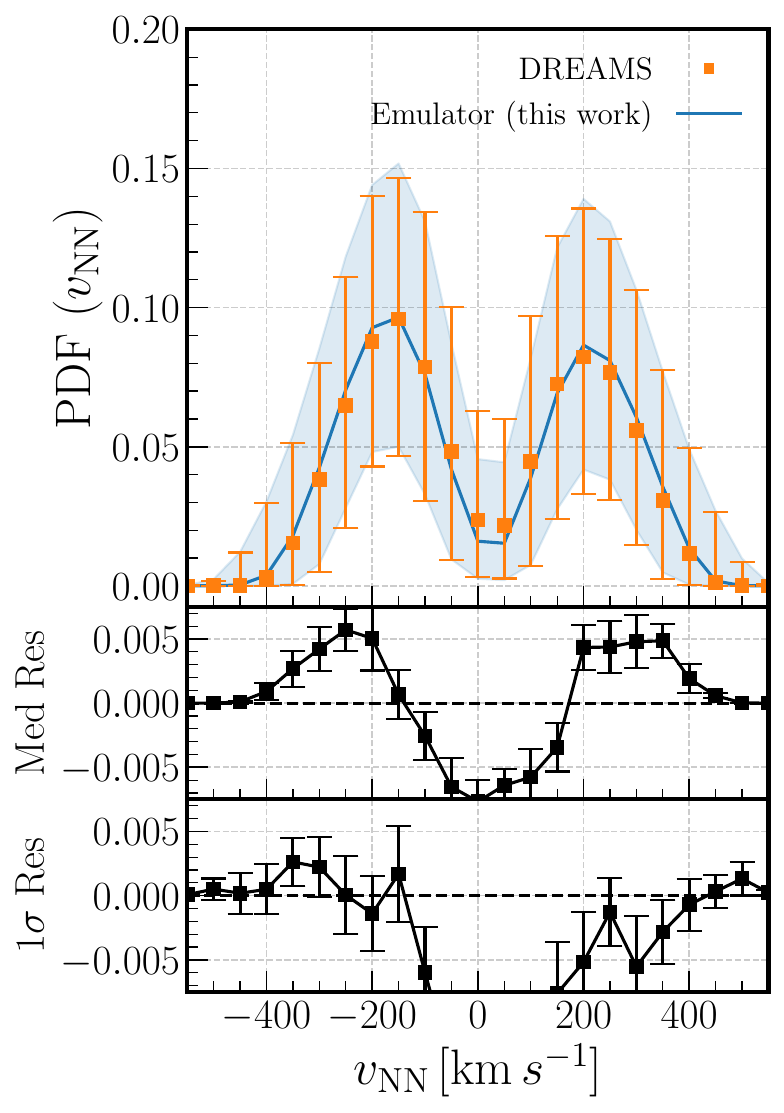}
    \caption{
    The distributions of radial velocities (left), pairwise velocities (middle), and the nearest-neighbor velocities (right).
    The top panels show the medians and $1\sigma$ intervals of the CDFs for the \nn samples (blue band) and the simulations (error bars).
    The middle and bottom panels show the residuals between the \nn samples and the simulations of the medians and the $1\sigma$ intervals, respectively.
    }
    \label{fig:v_pairwise_cdf}
\end{figure*}

We compute the distributions of radial velocity $\mathrm{v}_\mathrm{rad}$, the pairwise velocities $\mathrm{v}_\mathrm{pair}$, and the nearest-neighbor velocities $\mathrm{v}_\mathrm{NN}$.
For simplicity, the radial velocity of each satellite is calculated from the center of the \textit{halo} (using the reference frame described in Section~\ref{section:method}).
In other words, we do not take into account the velocity offsets between the halos and their central galaxies.
These offsets are typically much smaller compared to the velocity of the satellites (order of $10 \,\mathrm{km \, s^{-1}}$ compared to $100\,\mathrm{km \, s^{-1}}$ ), so we do not expect them to significantly affect our results.
The nearest-neighbor velocity of each satellite is determined by identifying its nearest neighbor (in position space) and calculating the velocity between them.
For each satellite, we assign its velocity a positive (negative) sign if it moves away from (towards) the central galaxy for $\mathrm{v}_\mathrm{rad}$, or another satellite satellite for $\mathrm{v}_\mathrm{pair}$ and $\mathrm{v}_\mathrm{NN}$.

Figure~\ref{fig:v_pairwise_cdf} shows the distributions of the radial velocities $\mathrm{v}_\mathrm{rad}$ (left), the pairwise velocities $\mathrm{v}_\mathrm{pair}$ (middle), and the nearest-neighbor velocities $\mathrm{v}_\mathrm{NN}$ (right) of satellites of the \nn samples (blue) and the simulations (orange).
Similar to Figure~\ref{fig:r_pairwise_cdf}, the medians, 16th, and 84th percentiles are shown. 
Note that, unlike the nearest-neighbor separations, we do not observe strong variations of the nearest-neighbor velocities with the simulation parameters, and thus do not show results for individual bins. 
For $\mathrm{v}_\mathrm{pair}$ and $\mathrm{v}_\mathrm{NN}$, we show their probability distribution functions (PDFs) instead CDFs as the comparison is easier to interpret with PDFs due to the bimodality in the distributions (described below).
Additionally, we note that although both peaks are at the same absolute velocity ($\pm 200 \, \mathrm{km\, s^{-1}}$, the negative peak is slightly more prominent, suggesting that the overall dynamics of the system have more infall.

Overall, \nn can predict the general trend in both the median and the $1\sigma$ percentiles of the velocity distributions.
However, we note similar discrepancies to those observed in the spatial distributions.
In particular, \nn underpredicts the distributions of satellites moving at $-100 \, \mathrm{km\,s^{-1}} < \mathrm{v}_\mathrm{pair}, \mathrm{v}_\mathrm{NN} < 100 \, \mathrm{km\,s^{-1}}$.
This implies that generated satellites are typically moving relatively to each other faster than those in the simulations.
Additionally, \nn also underpredicts the distributions of $ \mathrm{v}_\mathrm{pair}$ at high negative velocity, approximately $-400 \, \mathrm{km\,s^{-1}}$.
Since this occurs in the tail of the distribution, the discrepancy could be attributed to the limited number of training samples in this velocity range. 
Similarly, we observe a better agreement between the radial velocity $\mathrm{v}_\mathrm{rad}$ distributions of the \nn samples and simulations.
We discuss the limitations of the current model and potential ways to overcome them in Section~\ref{section:discussion}.


\label{lastpage}
\end{document}